\pdfoutput=1

\documentclass[iop]{emulateapj}

\slugcomment{}

\shorttitle{Observational signatures of cloud-cloud collision in the E-S235}
\shortauthors{L.~K. Dewangan}
\begin{document}

\title{Observational signatures of cloud-cloud collision in the extended star-forming region S235}
\author{L.~K. Dewangan\altaffilmark{1} and D.~K. Ojha\altaffilmark{2}}
\email{lokeshd@prl.res.in}
\altaffiltext{1}{Physical Research Laboratory, Navrangpura, Ahmedabad - 380 009, India.}
\altaffiltext{2}{Department of Astronomy and Astrophysics, Tata Institute of Fundamental Research, Homi Bhabha Road, Mumbai 400 005, India.}
\begin{abstract}
We present a multi-wavelength data analysis of the extended star-forming region S235 (hereafter E-S235), 
where two molecular clouds are present. 
In E-S235, using the $^{12}$CO (1-0) and $^{13}$CO (1-0) line data, a molecular cloud linked with the site ``S235main" is traced in a velocity range [$-$24, $-$18] km s$^{-1}$, 
while the other one containing the sites S235A, S235B, and S235C (hereafter ``S235ABC") is depicted in a velocity range [$-$18, $-$13] km s$^{-1}$.
In the velocity space, these two clouds are separated by $\sim$4 km s$^{-1}$, and are interconnected by a lower intensity intermediate 
velocity emission, tracing a broad bridge feature. In the velocity channel maps, a possible complementary molecular pair at [$-$21, $-$20] km s$^{-1}$ and [$-$16, $-$15] km s$^{-1}$ is also evident. 
The sites, ``S235ABC", East~1, and South-West are spatially seen in the interface of two clouds.
Together, these observed features are consistent with the predictions of numerical models of the cloud-cloud collision (CCC) process, favoring 
the onset of the CCC in E-S235 about 0.5 Myr ago. Deep UKIDSS near-infrared photometric analysis of point-like sources reveals significant clustering of young stellar populations toward 
 the sites located at the junction, and the ``S235main".
The sites, ``S235ABC" harbor young compact H\,{\sc ii} regions having dynamical 
 ages of $\sim$0.06--0.22 Myr, and these sites (including South-West and East~1) also contain dust clumps (having M$_{clump}$ $\sim$40 to 635 M$_{\odot}$). 
 Our observational findings suggest that the star formation activities (including massive stars) appear to be influenced by the CCC mechanism at the junction. 
 \end{abstract}
\section{Introduction}
\label{sec:intro}
The young stellar populations can be formed spontaneously throughout a given molecular cloud \citep[e.g.,][]{preibisch07}. 
On the other hand, the collapse and star formation can be influenced by some external agents such as the expansion of H\,{\sc ii} 
regions \citep[e.g.,][]{elmegreen77,elmegreen98} and the cloud-cloud collision (CCC) process \citep[e.g.,][]{habe92,anathpindika10,inoue13,takahira14}. 
In recent years, the study of the triggered star formation through the CCC process is an interesting and important issue in the star formation research \citep[e.g.,][and references therein]{torii17}. 
It has also been suggested that the CCC process can form massive OB stars and young stellar clusters at the junction of molecular clouds \citep[e.g.,][]{habe92,furukawa09,ohama10,inoue13,takahira14,fukui14,fukui16,torii15,torii17,haworth15a,haworth15b,dewangan17}.
In hydrodynamical numerical simulations, it has been shown that the CCC triggers the birth of dense clumps/cores in the shock-compressed interface, 
and these dense clumps/cores can further lead to the formation of a new generation 
of stars (\citet{habe92,anathpindika10}, and also see Figure~1 in \citet{torii17}). 
Furthermore, \citet{takahira14} studied the formation and evolution of pre-stellar gas cores in the collision of non-identical clouds with Bonner-Ebert profiles using hydrodynamical simulations. 
Recent magnetohydrodynamical numerical simulations have demonstrated that the CCC process can enable 
the formation of the massive clumps in the collisional-compressed layer \citep{inoue13}. 
More recently, based on the comparisons between observations and existing models, 
\citet{torii17} reported the characteristic features of the CCC, such as the complementary distribution of the two 
colliding clouds, the bridge feature at the intermediate velocity range, and its flattened CO spectrum. 
These features are promising to obtain convincing evidences to support the idea that two molecular components 
are interacting with each other in a given star-forming complex. 
\citet{torii17} also provided a table of promising star-forming sites 
where the star formation process could be explained by the CCC process.  
However, the investigation of observational signatures of star formation (including massive stars) via the CCC mechanism is still rare and 
very challenging. 

The extended star-forming region S235 (hereafter E-S235) is part of the giant molecular cloud G174+2.5 in the Perseus Spiral Arm \citep[e.g.][]{heyer96} and 
is a very well studied site using the multi-wavelength data covering from optical, near-infrared (NIR) to radio wavelengths \citep[e.g.,][etc.]{georgelin73,israel78,evans81a,evans81b,nordh84,brand93,felli97,felli04,felli06,allen05,kirsanova08,kirsanova14,boley09,dewangan11,
camargo11,kang12,chavarria14,navarete15,foster15,burns15,burns16,ladeyschikov16,bieging16,dewangan16}.
Located at a distance of 1.8 kpc \citep{evans81a}, E-S235 is a nearby star-forming region that consists of previously known star-forming sites, 
``S235 complex (or S235main)", S235A, S235B, and S235C (or ``S235ABC") (see Figure~1 in \citet{dewangan11} (hereafter Paper I) and also in \citet{kirsanova14}). 
In E-S235, \citet{evans81a} found two velocity components at $-$20 km s$^{-1}$ (linked with ``S235main") and $-$17 km s$^{-1}$ (containing the sites ``S235ABC"). 
This implies that the molecular gas linked with the sites ``S235ABC" is redshifted with respect to the S235main molecular cloud. 
Furthermore, using $^{13}$CO (1$-$0) line data, 
\citet{kirsanova08} traced three molecular gas components (i.e., $-18$ km s$^{-1}$ $< V_{lsr} < -15$ km s$^{-1}$ (red), $-21$ km s$^{-1}$ $< V_{lsr} < -18$ km s$^{-1}$ (central), and $-25$ km s$^{-1}$ $< V_{lsr} < -21$ km s$^{-1}$ (blue)) in the direction of E-S235. 
An extended H\,{\sc ii} region linked with the ``S235main" is ionized by a single massive star BD+35$^{o}$1201 of O9.5V type \citep{georgelin73}, 
while the sites, S235A, S235B, and S235C are associated with compact H\,{\sc ii} regions, which are mainly excited by B1V--B0.5V stars 
\citep[see Table~1 in][and references therein]{bieging16}. 
The existing observations and interpretations of E-S235 suggest that the H\,{\sc ii} regions associated with 
the ``S235main" and the ``S235ABC" are interacting with their surrounding molecular clouds. 
Photometric analysis of point-like sources indicated the intense star formation activities in these molecular clouds \citep[][]{kirsanova08,dewangan11,chavarria14,dewangan16,bieging16}.  
Furthermore, these sites have been cited as promising regions of triggered star formation by the expanding H\,{\sc ii} regions \citep{kirsanova08,kirsanova14,camargo11,bieging16}. 
\citet{dewangan16} (hereafter Paper II) performed a multi-wavelength study of star formation activity in the ``S235main" and 
found that at least five subregions (i.e. Central~E, East~2, North, North-West, and Central~W) (having A$_{V}$ $>$ 8 mag) 
appear to be nearly regularly spaced along the sphere-like shell surrounding the ionized emission. 
These results have also been seen in the molecular and dust continuum maps. 
They concluded that the ``S235main" can be considered as a promising nearby site of triggered star formation. 
Furthermore, in E-S235, they also found an almost broad bridge feature in the velocity space and 
suggested a possibility of the applicability of the CCC process. They suggested that the CCC process might have influenced the star formation activity 
at the interface between the S235main molecular cloud and the ``S235ABC" molecular 
cloud (i.e. South-West subregion). In Paper~II, the analysis was focused mainly on the ``S235main" site.
Hence, a detailed study of the CCC process concerning the formation of young stellar clusters and massive 
stars in E-S235 is yet to be carried out.

To explore the CCC process in E-S235, we have used the narrow-band H$\alpha$ image from the 
Isaac Newton Telescope Photometric H$\alpha$ Survey of the Northern Galactic Plane \citep[IPHAS;][]{drew05}, NIR data from the UKIDSS Galactic Plane Survey \citep[GPS;][]{lawrence07}, radio continuum data at 610 MHz from the Giant Metre-wave Radio Telescope (GMRT) database, and 
dust continuum 1.1 mm data \citep{aguirre11,ginsburg13} from the Bolocam Galactic Plane Survey (BGPS).
Additionally, several published data (i.e. {\it Spitzer} NIR and mid-infrared (MIR) 
data \citep[from][]{dewangan11}, and the Five College Radio Astronomy Observatory (FCRAO) $^{12}$CO (1-0) and $^{13}$CO (1-0) line data \citep{heyer98,brunt04}) were also utilized. 

This paper is structured as follows. In Section~\ref{sec:obser}, we discuss the data selection. 
The results of our extensive multi-wavelength data analysis are presented in Section~\ref{sec:data}. 
The implications of our findings concerning the star formation are discussed in Section~\ref{sec:disc}.
Finally, the results are summarized and concluded in Section~\ref{sec:conc}.
\section{Data and analysis}
\label{sec:obser}
In this work, we have selected a region of $\sim$24$\arcmin$ $\times$ 24$\arcmin$ (or $\sim$ 12.5 pc  $\times$ 12.5 pc at a distance of 1.8 kpc)
(central coordinates: $\alpha_{2000}$ = 05$^{h}$41$^{m}$02$^{s}$, 
$\delta_{2000}$ = +35$\degr$47$\arcmin$25$\arcsec$) containing E-S235.
In the following, we provide a brief description of the adopted multi-frequency data-sets. 
\subsection{Narrow-band H$\alpha$ image}
Narrow-band H$\alpha$ image at 0.6563 $\mu$m was retrieved from the IPHAS database. 
The survey was carried out using the Wide-Field Camera (WFC) at the 2.5-m Isaac Newton Telescope, located at La Palma. 
The WFC contains four 4k $\times$ 2k CCDs, in an L-shape configuration. The pixel
scale is $0\farcs33$ and the instantaneous field of view is about 0.3 square degrees. 
More details about the IPHAS can be found in \citet{drew05}. 
\subsection{Near-infrared data}
We analyzed the deep NIR photometric {\it HK} magnitudes of point sources extracted from the UKIDSS GPS sixth archival data release (UKIDSSDR6plus) and Two Micron All Sky Survey \citep[2MASS;][]{skrutskie06}. 
The UKIDSS observations (resolution $\sim$$0\farcs8$) were performed with the WFCAM mounted on the United Kingdom Infra-Red Telescope. 
2MASS photometric data were utilized to calibrate the final fluxes. 
In this work, we extracted only a reliable NIR photometric catalog. 
More information about the selection procedures of the GPS photometry can be found in \citet{dewangan15}.
Our resultant GPS catalog consists of point sources fainter than H = 12.3 and K = 11.4 mag to avoid saturation. 
In our final catalog, the magnitudes of the saturated bright sources were obtained from the 2MASS catalog. 

Additionally, we also obtained {\it Spitzer} 3.6 $\mu$m image from Paper~I.
\subsection{Dust continuum 1.1 mm data}
Bolocam 1.1 mm image and Bolocam source catalog (v2.1) at 1.1 mm were extracted from the BGPS. 
The effective full width at half maximum (FWHM) of the 1.1 mm map is $\sim$33$\arcsec$. 
\subsection{Radio continuum data}
The  610 MHz continuum data were obtained from the GMRT archive and were observed on 18--19 June 2005 (Project Code: 08SKG01). 
The data reduction was performed using AIPS software, in a manner similar to that highlighted in \citet{mallick13}. 
Since the S235 complex was distant from the center of the observed field, 
the primary beam correction for 610 MHz was also done, using the AIPS PBCOR task and parameters 
from the GMRT manual\footnote[1]{http://gmrt.ncra.tifr.res.in/gmrt\_hpage/Users/doc/obs\_manual.pdf}. 
The synthesized beam size of the final 610 MHz map is $\sim$48$\arcsec$ $\times$ 44\farcs2 (also see Paper~II). 
\subsection{Molecular CO line data}
In E-S235, we probed the molecular gas content using the FCRAO $^{12}$CO (J=1$-$0) and $^{13}$CO (J=1$-$0) line data. 
The FCRAO beam sizes are 45$\arcsec$ and 46$\arcsec$ for $^{12}$CO and $^{13}$CO, respectively. 
The observations of E-S235 were made as part of the Extended Outer Galaxy Survey \citep[E-OGS,][]{brunt04}, 
that extends the coverage of the FCRAO Outer Galaxy Survey \citep[OGS,][]{heyer98} to Galactic longitude $l$ = 193$\degr$, 
over a latitude range of $-$3$\degr$.5 $\leq$ $b$ $\leq$ +5$\degr$.5. 
However, the data cubes of E-S235 were further re-processed and a document describing the re-processing 
data methods is given in \citet{brunt04} (also Brunt C.M. et al.; in preparation).
These $^{13}$CO data cubes were collected from M. Heyer and C. Brunt (through private communication). 
The FCRAO $^{13}$CO (J=1$-$0) line data were also used in Paper~II.
\section{Results}
\label{sec:data}
In this section, we present a multi-wavelength analysis of E-S235 to understand the ongoing physical processes operating in this target region. 
\subsection{Kinematics of molecular gas in E-S235}
\label{sec:coem} 
\begin{figure*}
\epsscale{0.9}
\plotone{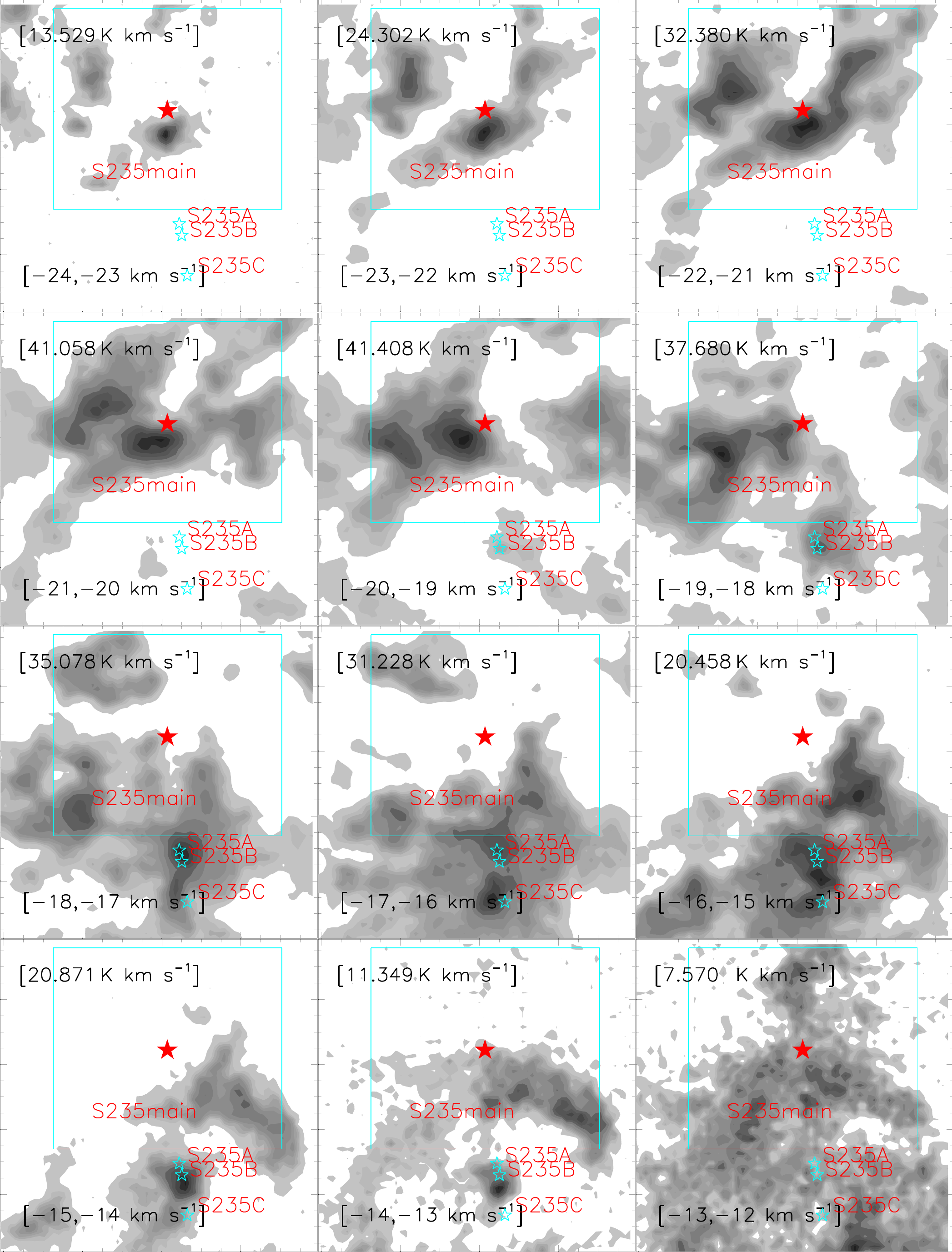}
\caption{\scriptsize The $^{12}$CO(J =1$-$0) velocity channel contour maps.
The molecular emission is integrated over a velocity interval, which is given in each panel (in km s$^{-1}$). 
The contour levels are 10, 20, 25, 30, 35, 40, 50, 60, 70, 80, 90, and 98\% of the peak value (in K km s$^{-1}$), 
which is also provided in each panel. The positions of ionizing stars of S235A, S235B, and S235C are marked by open stars. 
In the ``S235main", the location of an O9.5V star (BD+35$^{\degr}$1201) is highlighted by a filled red star. 
The solid box shows the area investigated in Paper~II which is related to the site ``S235main".}
\label{fig2}
\end{figure*}

\begin{figure*}
\epsscale{0.9}
\plotone{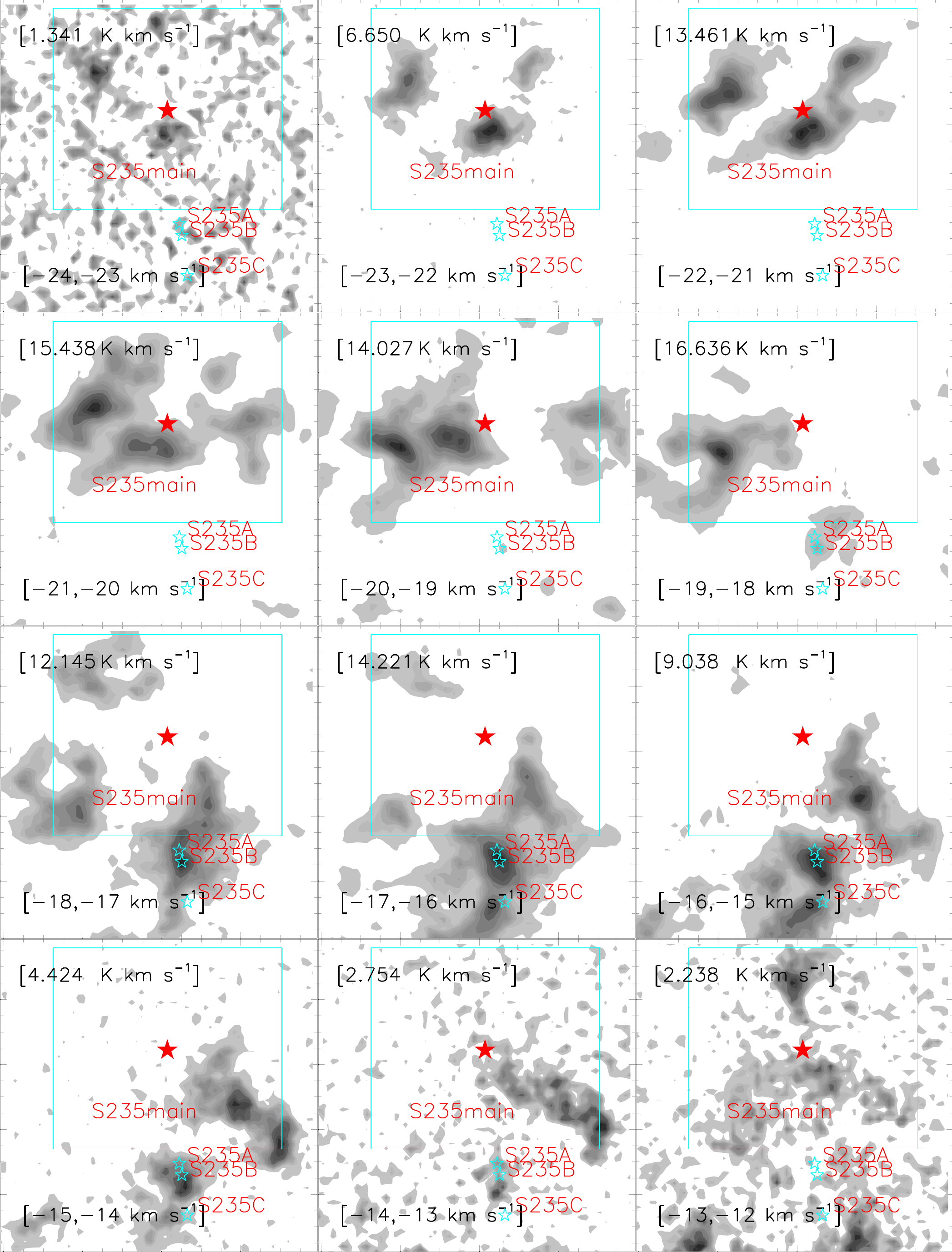}
\caption{\scriptsize The $^{13}$CO(J =1$-$0) velocity channel contour maps.
The molecular emission is integrated over a velocity interval, which is given in each panel (in km s$^{-1}$). 
The contour levels are 10, 20, 25, 30, 35, 40, 50, 60, 70, 80, 90, and 98\% of the peak value (in K km s$^{-1}$), 
which is also provided in each panel. In each panel, other marked symbols and labels are similar to those shown in Figure~\ref{fig2}.}
\label{fig3}
\end{figure*}

\begin{figure*}
\epsscale{1}
\plotone{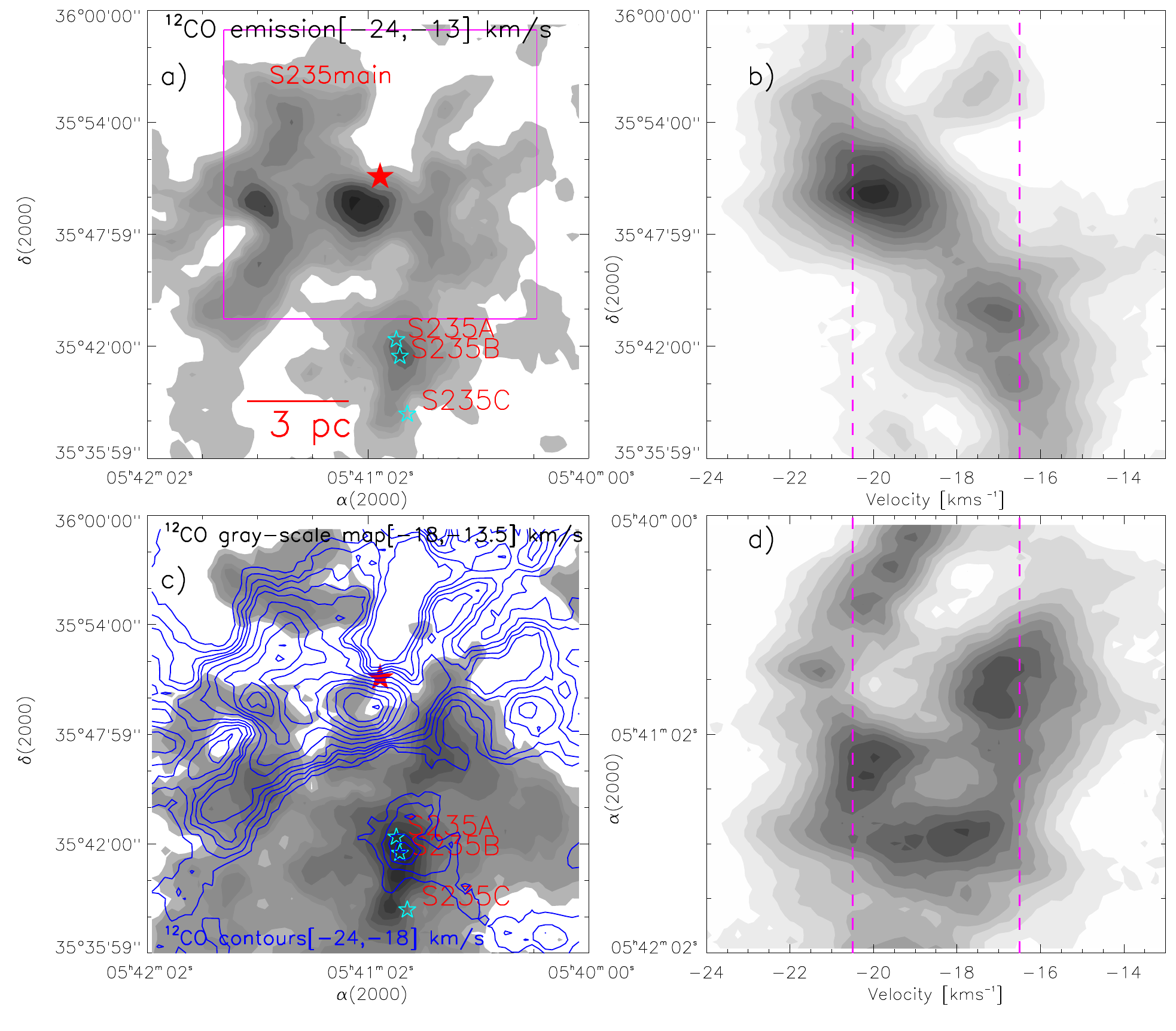}
\caption{\scriptsize The distribution of molecular gas toward E-S235 (including ``S235main" and ``S235ABC").
a) Integrated intensity map of $^{12}$CO (J = 1-0) from $-$24 to $-$13 km s$^{-1}$. 
The contour levels are 20, 30, 35, 40, 50, 60, 70, 80, 90, and 98\% of the 
peak value (i.e. 130.929 K km s$^{-1}$). b) Declination-velocity map of $^{12}$CO. 
The $^{12}$CO emission is integrated over the right ascension range from 05$^{h}$42$^{m}$02$^{s}$ to 05$^{h}$40$^{m}$00$^{s}$.
c) Two molecular components (a redshifted and a blueshifted) in the direction of E-S235. 
The $^{12}$CO emission contours from $-24$ to $-$18 km s$^{-1}$ are overplotted on the 
$^{12}$CO emission map. 
The background $^{12}$CO emission map (from $-$18 to $-$13.5 km s$^{-1}$) is shown with levels of 
79.628 K km s$^{-1}$ $\times$ (0.1, 0.15, 0.25, 0.3, 0.4, 0.5, 0.6, 0.7, 0.8, 0.9, and 0.98). 
The $^{12}$CO contours (in blue) are shown with levels of 
121.6 K km s$^{-1}$ $\times$ (0.1, 0.15, 0.2, 0.3, 0.45, 0.5, 0.6, 0.7, 0.8, 0.9, and 0.98). 
d) Right Ascension-velocity map of $^{12}$CO. 
The $^{12}$CO emission is integrated over the declination range from +35$\degr$35$\arcmin$59$\arcsec$ to 
+36$\degr$00$\arcmin$00$\arcsec$. 
In both the left panels (i.e. Figures~\ref{fig4}a and~\ref{fig4}c), the marked symbols are similar to those shown in Figure~\ref{fig2}. 
There are two velocity peaks (a blueshifted at $\sim-$20.5 km s$^{-1}$ 
and a redshifted at $\sim-$16.5 km s$^{-1}$) seen in the position-velocity 
maps (i.e. Figures~\ref{fig4}b and~\ref{fig4}d) which are highlighted by dashed magenta lines. These peaks are 
separated by a lower intensity intermediate velocity emission (i.e., a broad bridge feature; also see the text).}
\label{fig4}
\end{figure*}

The study of molecular gas in a given star-forming complex can help to trace the physical association of different subregions.
As mentioned earlier, two velocity components (at $-$20 and $-$17 km s$^{-1}$) are seen in the direction of E-S235 \citep{evans81a}. To explore the spatial distribution of molecular gas in E-S235, we present $^{12}$CO (J=1--0) and $^{13}$CO (J=1--0) velocity channel maps at different velocities covering a range from $-$24 to $-$12 km s$^{-1}$ in 
steps of 1 km s$^{-1}$ in Figures~\ref{fig2} and~\ref{fig3}, respectively. The analysis of channel maps of both $^{12}$CO and $^{13}$CO emission reveals 
that the molecular cloud associated with the site ``S235main" is traced in a velocity range of 
$-$23 to $-$18 km~s$^{-1}$, while the molecular cloud linked with the sites ``S235ABC" is depicted in a velocity range of $-$18 to $-$13 km~s$^{-1}$.
Together, our analysis is in agreement with the previously reported fact that the S235main molecular cloud is blueshifted with respect to 
the molecular cloud harboring the sites ``S235ABC". To further examine the velocity field and kinematical structure of the molecular gas in E-S235, using the $^{12}$CO line data, we present the integrated intensity map and the position-velocity maps in Figure~\ref{fig4}. In Figure~\ref{fig4}a, the $^{12}$CO emission is integrated over $-$24 to $-$13 km s$^{-1}$, tracing the gas distribution toward the sites, ``S235main" and ``S235ABC". In Figures~\ref{fig4}b and~\ref{fig4}d, we present the position-velocity diagrams of $^{12}$CO emission, 
tracing two molecular components as well as noticeable velocity gradients.  
In both the position-velocity maps, a redshifted molecular component (velocity peak at $\sim$$-$16.5 km s$^{-1}$) and a blueshifted component 
(velocity peak at $\sim$$-$20.5 km s$^{-1}$) are evident which are consistent with those observed in the channel maps of $^{12}$CO emission. 
This implies that these peaks are separated by $\sim$4 km s$^{-1}$. 
Furthermore, in Figure~\ref{fig4}b, we find that these two velocity peaks are interconnected by a lower intensity 
intermediate velocity emission, suggesting the presence of a broad bridge feature. 
In Figure~\ref{fig4}c, we present the spatial distribution of $^{12}$CO gas associated with the redshifted and the blueshifted molecular 
components within E-S235. 

\begin{figure*}
\epsscale{1}
\plotone{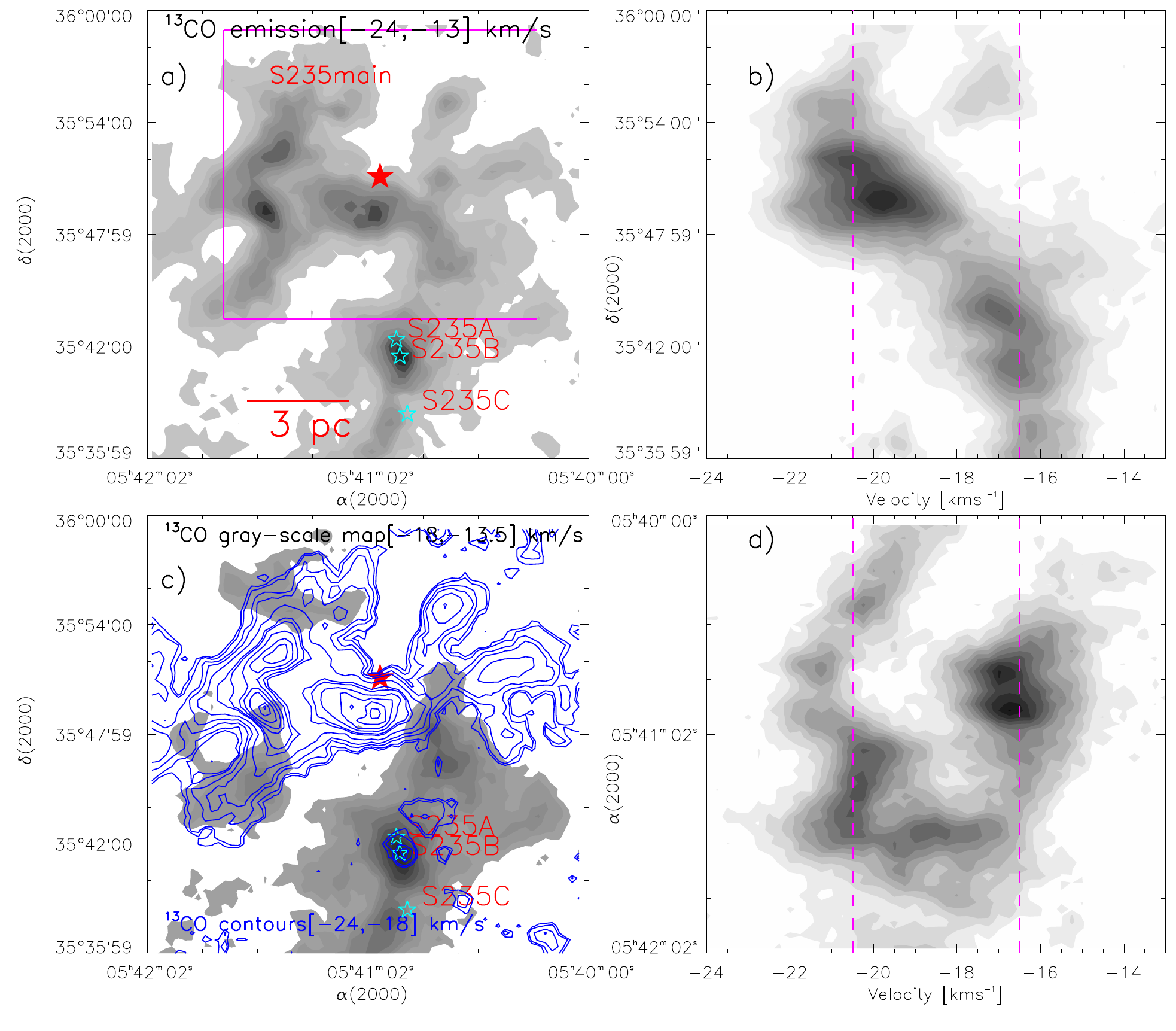}
\caption{\scriptsize The distribution of molecular gas toward E-S235 (including ``S235main" and ``S235ABC").
a) Integrated intensity map of $^{13}$CO (J = 1-0) from $-$24 to $-$13 km s$^{-1}$. 
The contour levels are 10, 20, 30, 35, 40, 50, 60, 70, 80, 90, and 98\% of the 
peak value (i.e. 36.046 K km s$^{-1}$). b) Declination-velocity map of $^{13}$CO. 
The $^{13}$CO emission is integrated over the right ascension range from 05$^{h}$42$^{m}$02$^{s}$ to 05$^{h}$40$^{m}$00$^{s}$.
c) Two molecular components (a redshifted and a blueshifted) in the direction of E-S235. 
The $^{13}$CO emission contours from $-24$ to $-$18 km s$^{-1}$ are overplotted on the 
$^{13}$CO emission map. 
The background $^{13}$CO emission map (from $-$18 to $-$13.5 km s$^{-1}$) is shown with levels of 
31.587 K km s$^{-1}$ $\times$ (0.09, 0.15, 0.25, 0.3, 0.4, 0.5, 0.6, 0.7, 0.8, 0.9, and 0.98). 
The $^{13}$CO contours (in blue) are shown with levels of 
32.252 K km s$^{-1}$ $\times$ (0.08, 0.1, 0.15, 0.2, 0.3, 0.45, 0.5, 0.6, 0.7, 0.8, 0.9, and 0.98). 
d) Right Ascension-velocity map of $^{13}$CO. 
The $^{13}$CO emission is integrated over the declination range from +35$\degr$35$\arcmin$59$\arcsec$ to 
+36$\degr$00$\arcmin$00$\arcsec$. 
In both the left panels (i.e. Figures~\ref{fig5}a and~\ref{fig5}c), the marked symbols are similar to those shown in Figure~\ref{fig2}. 
There are two velocity peaks (a blueshifted at $\sim-$20.5 km s$^{-1}$ 
and a redshifted at $\sim-$16.5 km s$^{-1}$) seen in the position-velocity 
maps (i.e. Figures~\ref{fig5}b and~\ref{fig5}d) which are highlighted by dashed magenta lines. These peaks are 
separated by a lower intensity intermediate velocity emission (i.e., a broad bridge feature; also see the text).}
\label{fig5}
\end{figure*}

\begin{figure*}
\epsscale{1}
\plotone{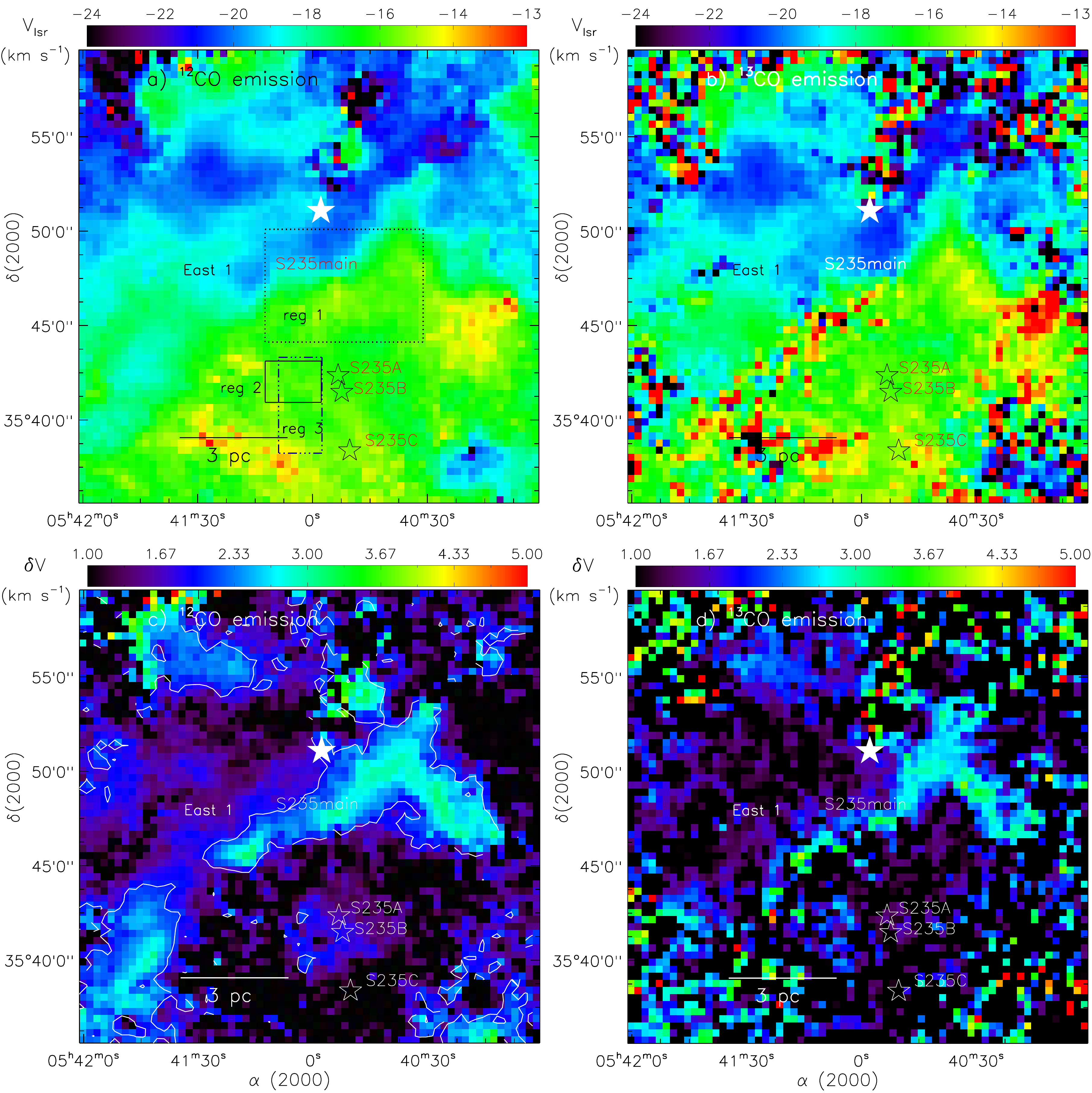}
\caption{\scriptsize a) $^{12}$CO first momentum map. 
Three small areas (i.e. reg 1, reg 2, and reg 3) are also highlighted by boxes in the figure. 
b) $^{13}$CO first momentum map. 
c) $^{12}$CO second momentum map (i.e. velocity dispersion ($\delta$V)). 
A velocity dispersion contour (in white) is also shown with a level of 2 km s$^{-1}$. 
d) $^{13}$CO second momentum map. 
In each top panel, the bar at the top shows the color-coded V$_{lsr}$ in km s$^{-1}$ (see Figures~\ref{fg5x}a and~\ref{fg5x}b).
In each bottom panel, the bar at the top shows the color-coded $\delta$V in km s$^{-1}$ (see Figures~\ref{fg5x}c and~\ref{fg5x}d).
In all the panels, the marked symbols are similar to those shown in Figure~\ref{fig4}a.}
\label{fg5x}
\end{figure*}

Using the $^{13}$CO line data, we also show the integrated intensity map and the position-velocity maps in Figure~\ref{fig5}.
Figure~\ref{fig5}a shows the $^{13}$CO intensity map integrated over $-$24 to $-$13 km s$^{-1}$.
Figures~\ref{fig5}b and~\ref{fig5}d present the position-velocity diagrams of $^{13}$CO emission, depicting also two molecular 
components as well as the broad bridge feature. 
We have found that the bridge feature is more noticeable in the velocity space of $^{12}$CO compared to $^{13}$CO data. 
Note that in Paper~II, only the declination-velocity map of $^{13}$CO related to E-S235 was discussed. 
In Paper~II, the presence of an expanding H\,{\sc ii} region was reported in ``S235main", 
on the basis of the detection of arc or ring-like features in the velocity space. 
Hence, in this work, we do not focus on these features seen in the velocity space. 
Figure~\ref{fig5}c presents the spatial distribution of $^{13}$CO gas linked with the blueshifted and 
the redshifted molecular components within E-S235. 
The distribution of $^{12}$CO and $^{13}$CO gas associated with both the clouds is spatially seen 
toward the sites, ``S235ABC" and a few subregions located in the East and South-West with respect 
to the location of the O9.5V star (BD+35$^{\degr}$1201) (see Figures~\ref{fig4}c and~\ref{fig5}c).
It implies the presence of junction/intersection zones in both the space and velocity between two molecular clouds.  
To further explore the boundaries of these two clouds, 
we present $^{12}$CO and $^{13}$CO first moment maps in Figures~\ref{fg5x}a and~\ref{fg5x}b, respectively.
These maps allow to reveal the V$_{lsr}$ of the peak emission at each grid point and are measure of the intensity-weighted mean velocity of the emitting gas. 
These maps clearly trace two distinct molecular clouds and also depict their boundaries in E-S235. 
Figures~\ref{fg5x}c and~\ref{fg5x}d show maps of $^{12}$CO and $^{13}$CO velocity dispersions ($\delta$V) 
(i.e. the second-order moment), enabling to infer the line width at each pixel. 
A large dispersion may be explained by a broad single velocity component 
and/or may indicate the presence of two or more narrow components with different velocities along one line of sight. 
In Figures~\ref{fg5x}c and~\ref{fg5x}d, the distribution of velocity dispersion has also enabled us to  
trace the boundaries of the molecular clouds or the colliding interface (where $\delta$V $>$ 2 km s$^{-1}$; 
see a white contour in Figure~\ref{fg5x}c). 

\begin{figure*}
\epsscale{0.44}
\plotone{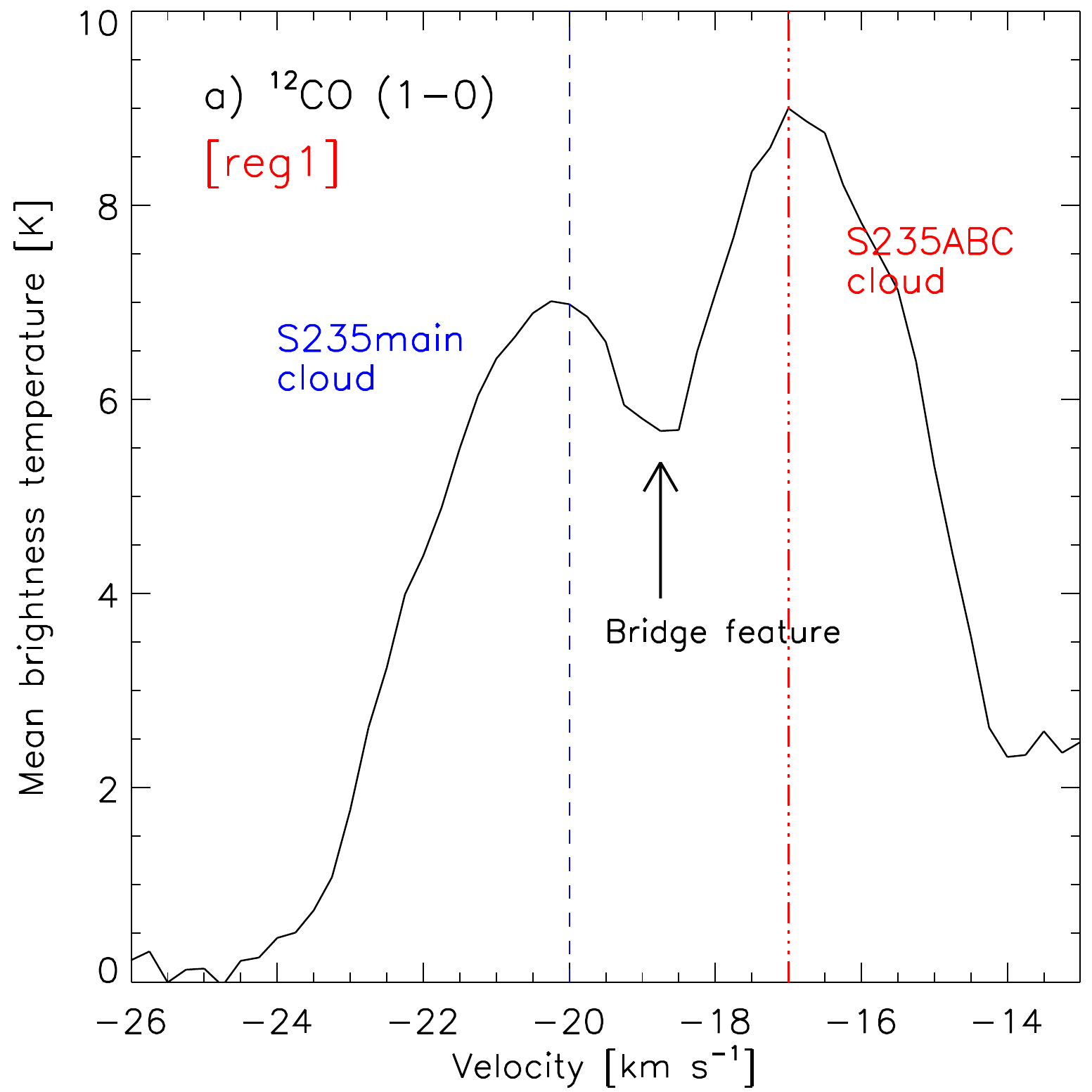}
\epsscale{0.44}
\plotone{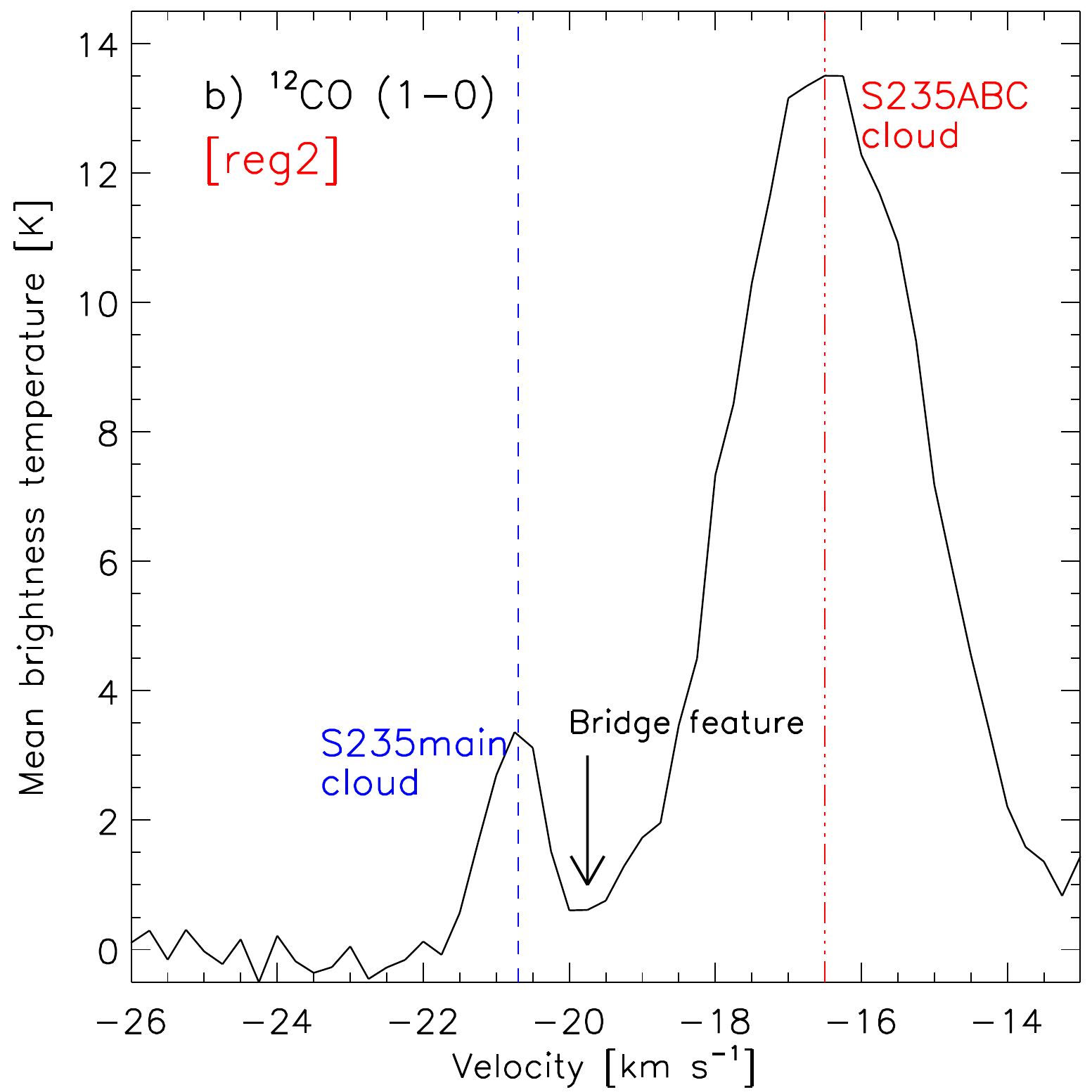}
\epsscale{0.44}
\plotone{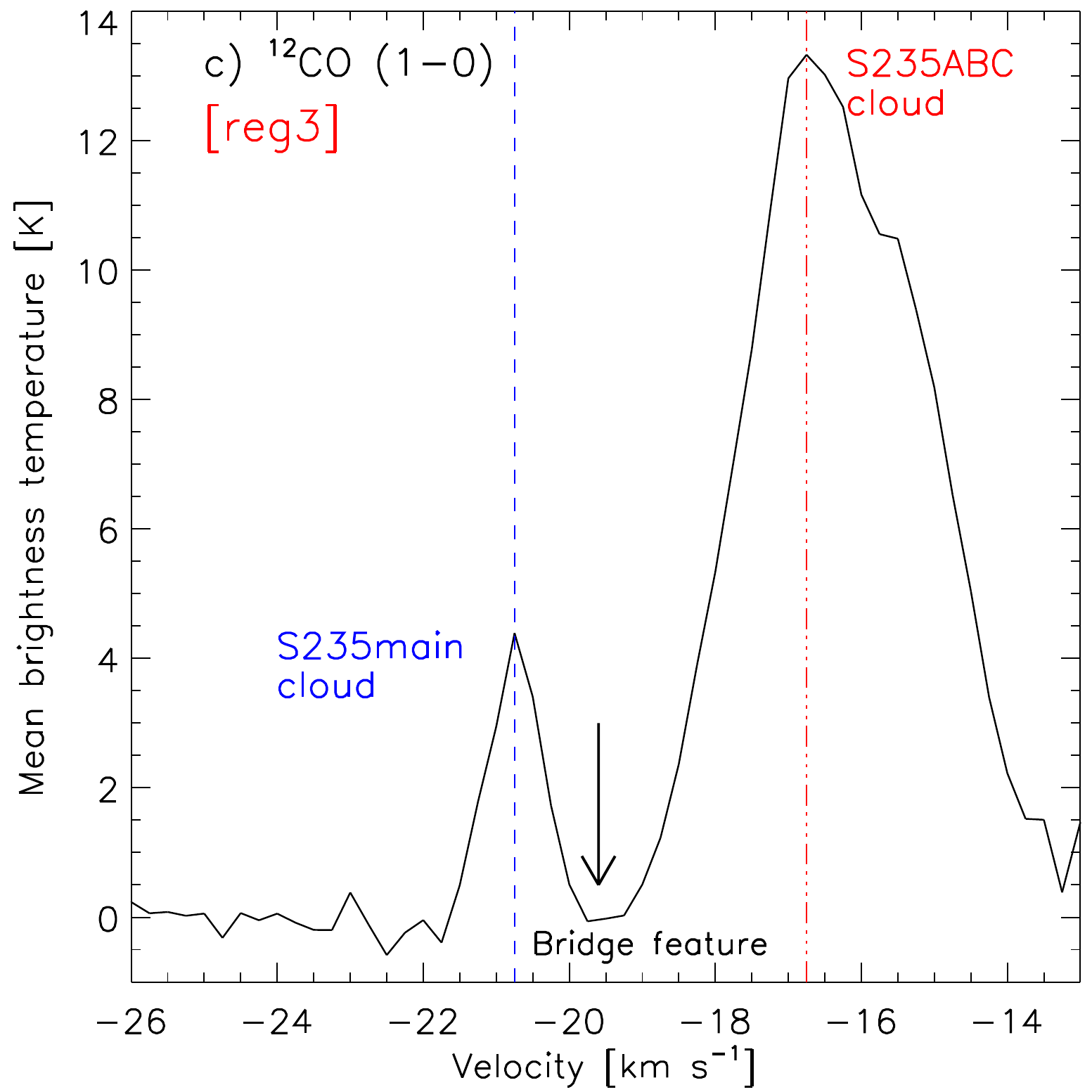}
\caption{\scriptsize $^{12}$CO (1-0) spectra toward our selected three small areas (i.e. reg 1, reg 2, and reg 3; see highlighted boxes in Figure~\ref{fg5x}a). 
Each spectrum is obtained by averaging each area. 
In each spectrum, an almost flattened profile is seen between two velocity peaks (i.e. a bridge feature at the intermediate velocity range).}
\label{fg6x}
\end{figure*}

We have further examined the broad bridge feature in E-S235 that contains H\,{\sc ii} regions. 
There are many physical processes (such as radiative/mechanical feedback from massive stars) which might act to remove the broad bridge feature in the vicinity of the H\,{\sc ii} regions.
Hence, it seems reasonable to look for the broad bridge features away from the H\,{\sc ii} regions. 
Figures~\ref{fg6x}a,~\ref{fg6x}b, and~\ref{fg6x}c show the observed $^{12}$CO (J=1--0) profiles toward three small areas, reg 1, reg2, and reg 3, respectively (see boxes in Figure~\ref{fg5x}a). 
Note that the area, reg 1 is chosen near the boundary of two molecular clouds, 
while the areas reg 2 and reg 3 are selected near the sites ``S235AB" and ``S235main", respectively.
Each spectrum is obtained by averaging the area highlighted by the respective box in Figure~\ref{fg5x}a.
Each spectrum shows the previously known two molecular clouds and also reveals a bridge feature seen at the intermediate velocity range that has almost flattened profile. In the star-forming region M20, \citet{torii17} also observed a similar feature in the $^{12}$CO spectra (see their Figure~14).

In the velocity space, the existence of a broad bridge feature indicates an observational signature of collisions 
between molecular clouds \citep[i.e. CCC process;][]{haworth15a,haworth15b}. 
\citet{torii17} pointed out that the bridge feature at the intermediate velocity range probes the turbulent motion of the gas enhanced by the collision. 
They also indicated that the spatially complementary distribution between two clouds is an expected outcome of the CCC and is attributed to identical silhouettes of the smaller cloud and the cavity in the larger cloud projected on the sky \citep[see Figure~1 in][]{torii17}. 
Based on this argument, we have also examined the channel maps of $^{12}$CO and $^{13}$CO, and have found a possible complementary pair that appears 
to be seen in the velocity ranges [$-$21, $-$20] km s$^{-1}$ and [$-$16, $-$15] km s$^{-1}$ (see Figures~\ref{fig2} and~\ref{fig3}). 

\begin{figure*}
\epsscale{0.5}
\plotone{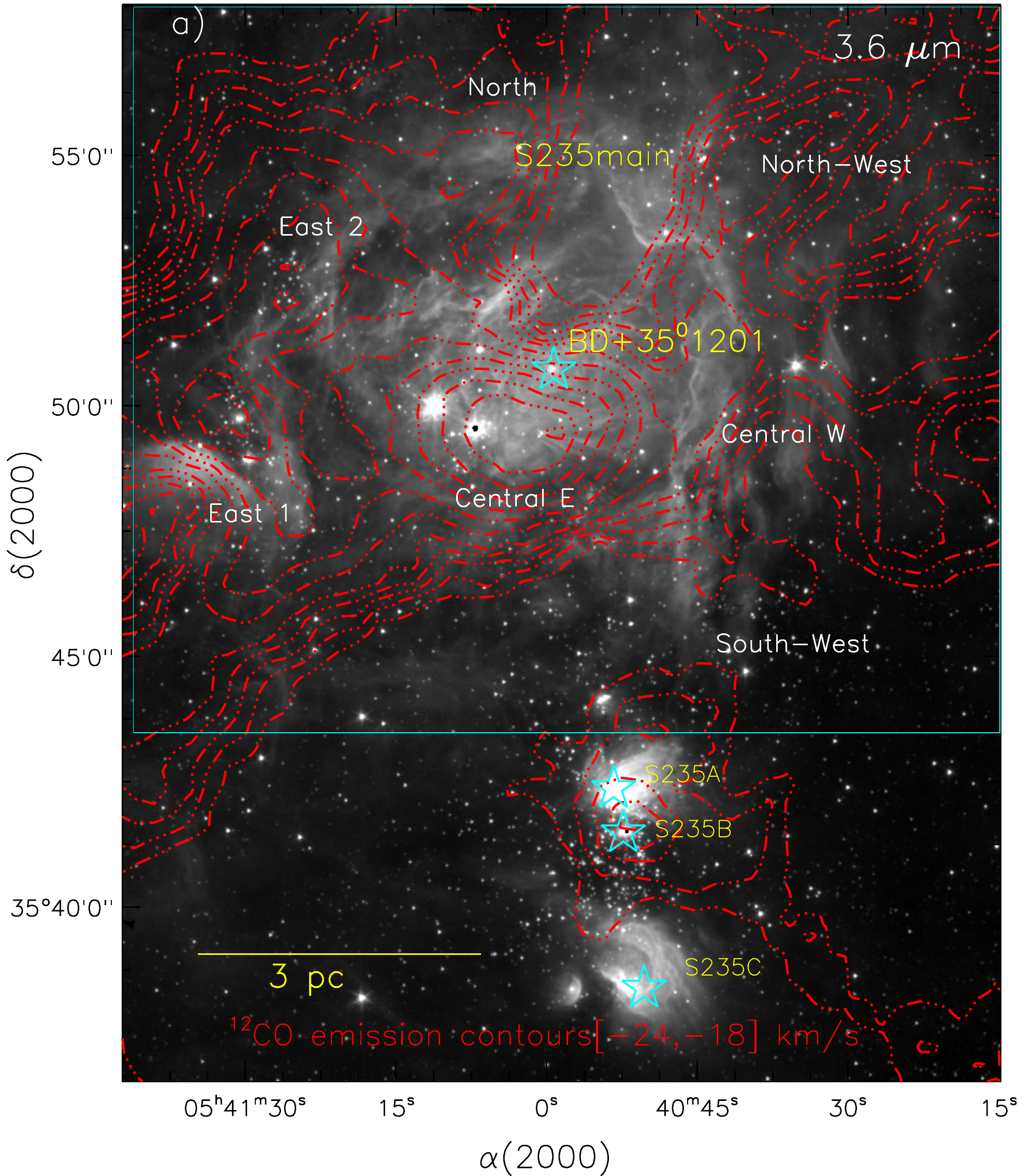}
\epsscale{0.5}
\plotone{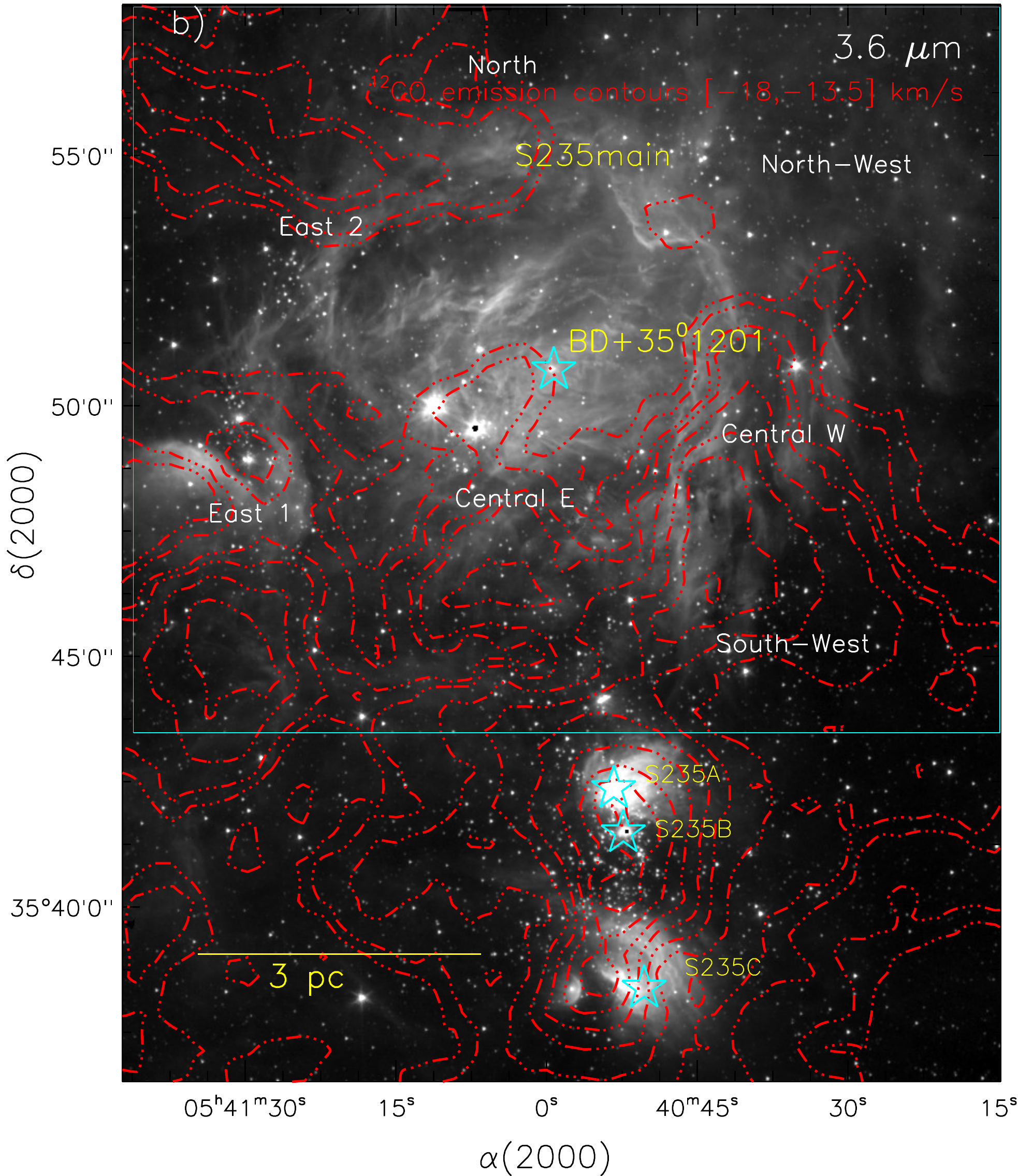}
\caption{\scriptsize a) {\it Spitzer} 3.6 $\mu$m image is overlaid with the $^{12}$CO emission integrated over $-$24 to $-$18 km s$^{-1}$. 
b) {\it Spitzer} 3.6 $\mu$m image is overlaid with the $^{12}$CO emission integrated over $-$18 to $-$13.5 km s$^{-1}$. 
In both the panels, different subregions are labeled \citep[also see Figure~1 in][]{dewangan16}. 
The scale bar on the bottom left shows a size of 3 pc at a distance of 1.8 kpc in each panel. 
In each panel, the contours are similar to the one shown in Figure~\ref{fig4}c. 
In both the panels, the positions of ionizing stars of ``S235main", S235A, S235B, and S235C are also highlighted by 
open stars \citep[also see Table~1 in][]{bieging16}. 
The positions of ionizing stars of S235A, S235B, and S235C are marked by open stars. 
In each panel, a solid box shows the area investigated in Paper~II.}
\label{fig6} 
\end{figure*}

\begin{figure*}
\epsscale{0.5}
\plotone{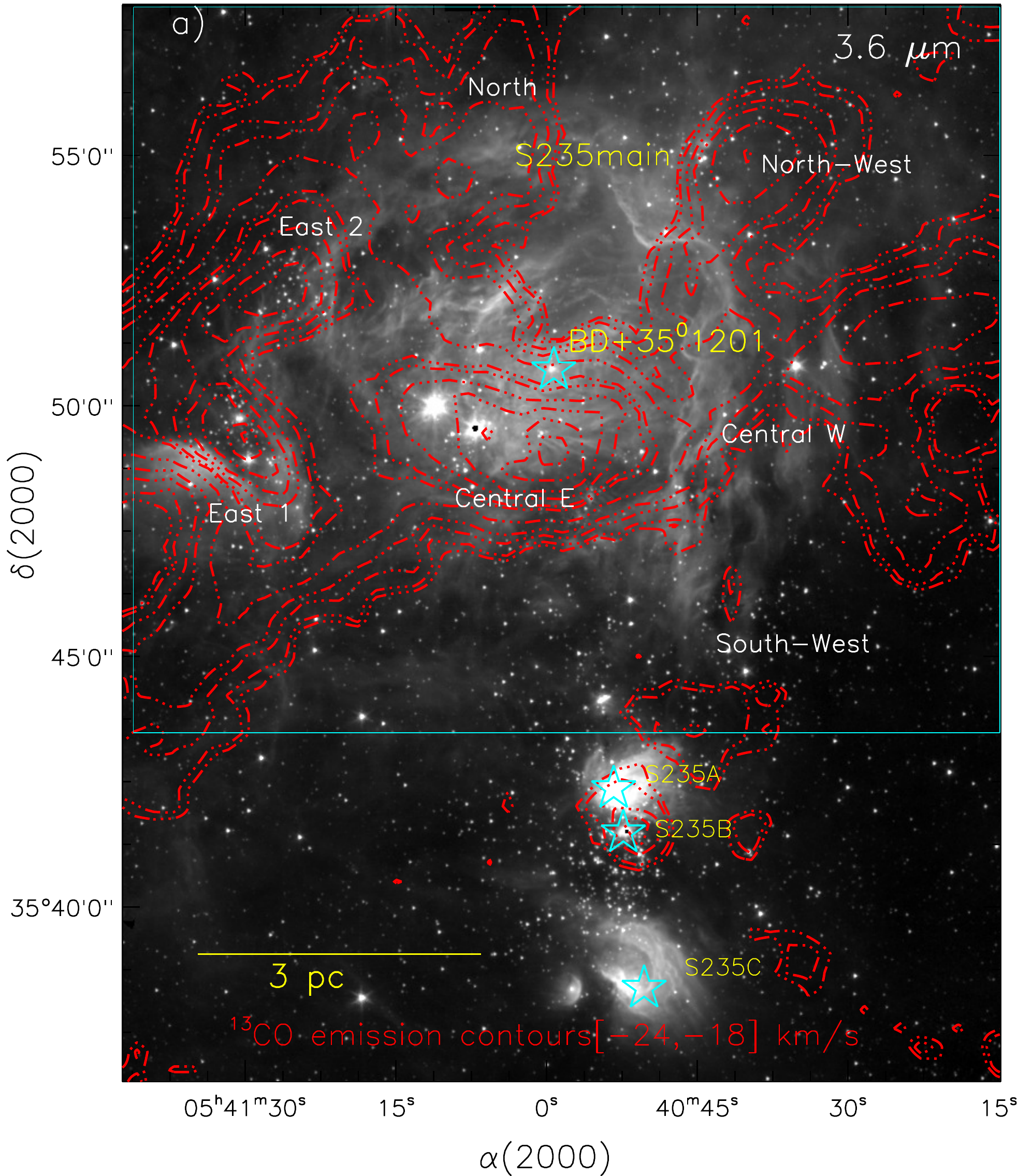}
\epsscale{0.5}
\plotone{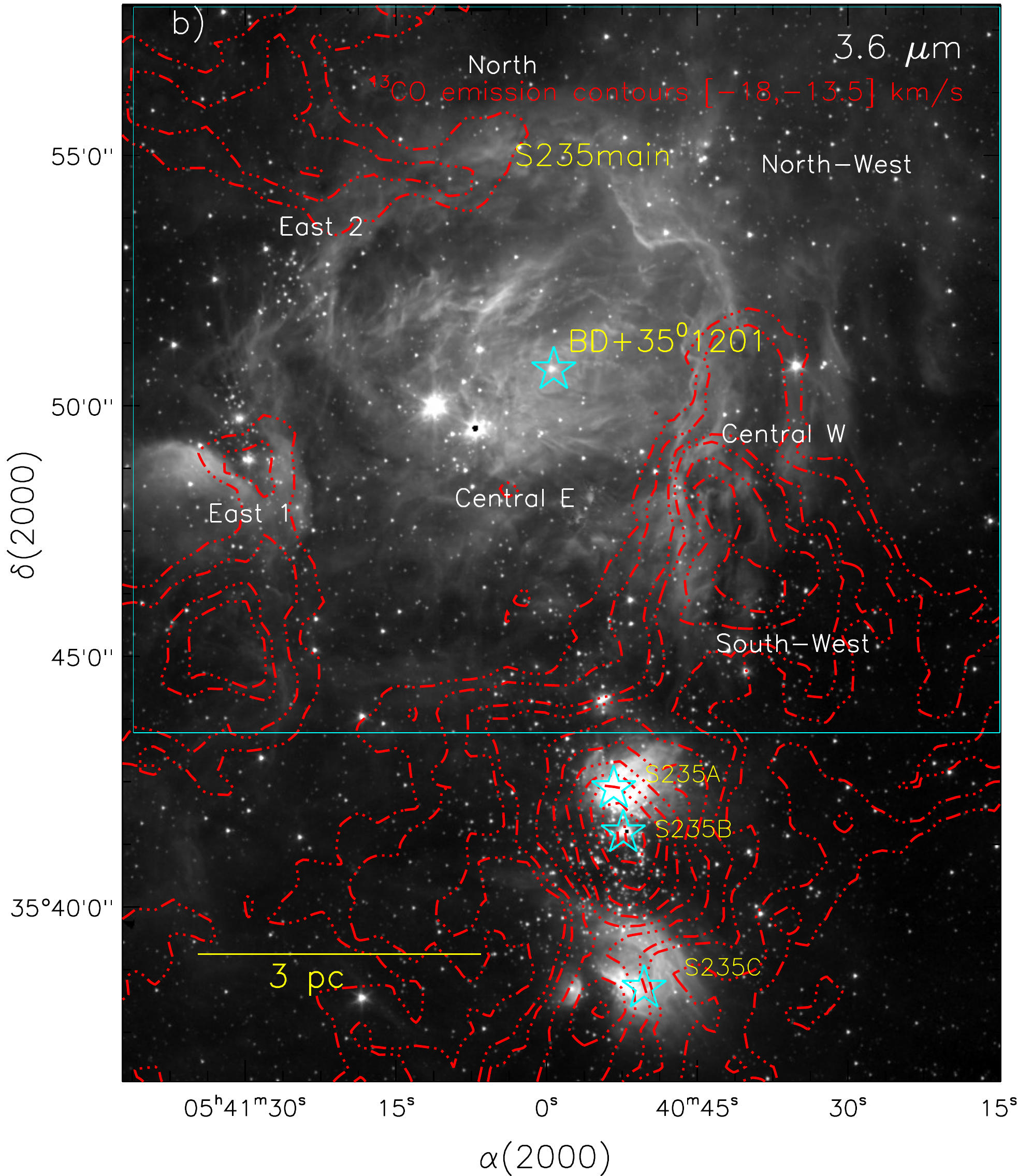}
\caption{\scriptsize a) {\it Spitzer} 3.6 $\mu$m image is overlaid with the $^{13}$CO emission integrated over $-$24 to $-$18 km s$^{-1}$. 
b) {\it Spitzer} 3.6 $\mu$m image is overlaid with the $^{13}$CO emission integrated over $-$18 to $-$13.5 km s$^{-1}$. 
In both the panels, different subregions are labeled \citep[also see Figure~1 in][]{dewangan16}. 
The scale bar on the bottom left shows a size of 3 pc at a distance of 1.8 kpc in each panel. 
In each panel, the contours are similar to the one shown in Figure~\ref{fig5}c. 
In each panel, other marked symbols and labels are similar to those shown in Figure~\ref{fig6}a.}
\label{fig7} 
\end{figure*}

\begin{figure*}
\epsscale{0.5}
\plotone{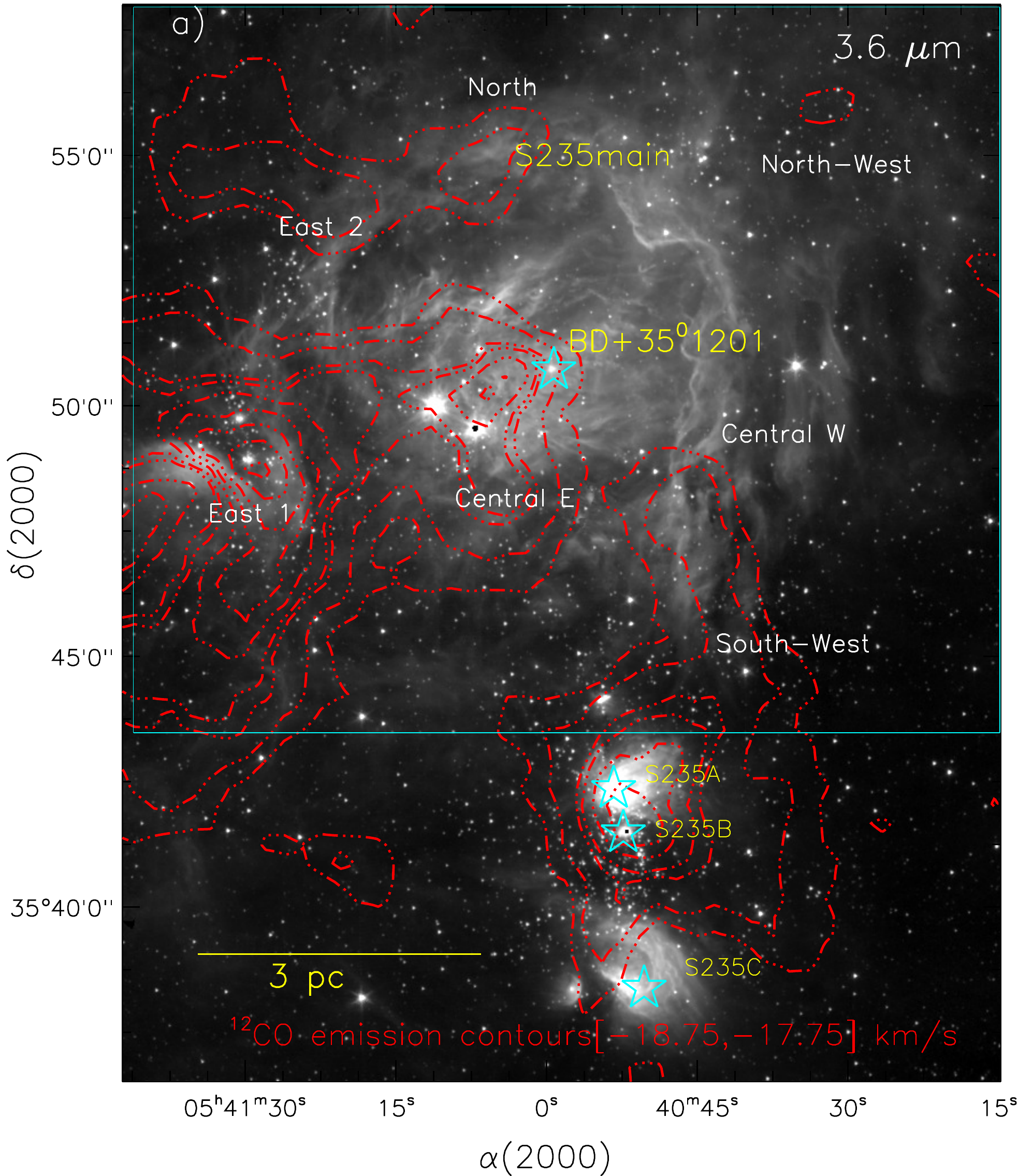}
\epsscale{0.5}
\plotone{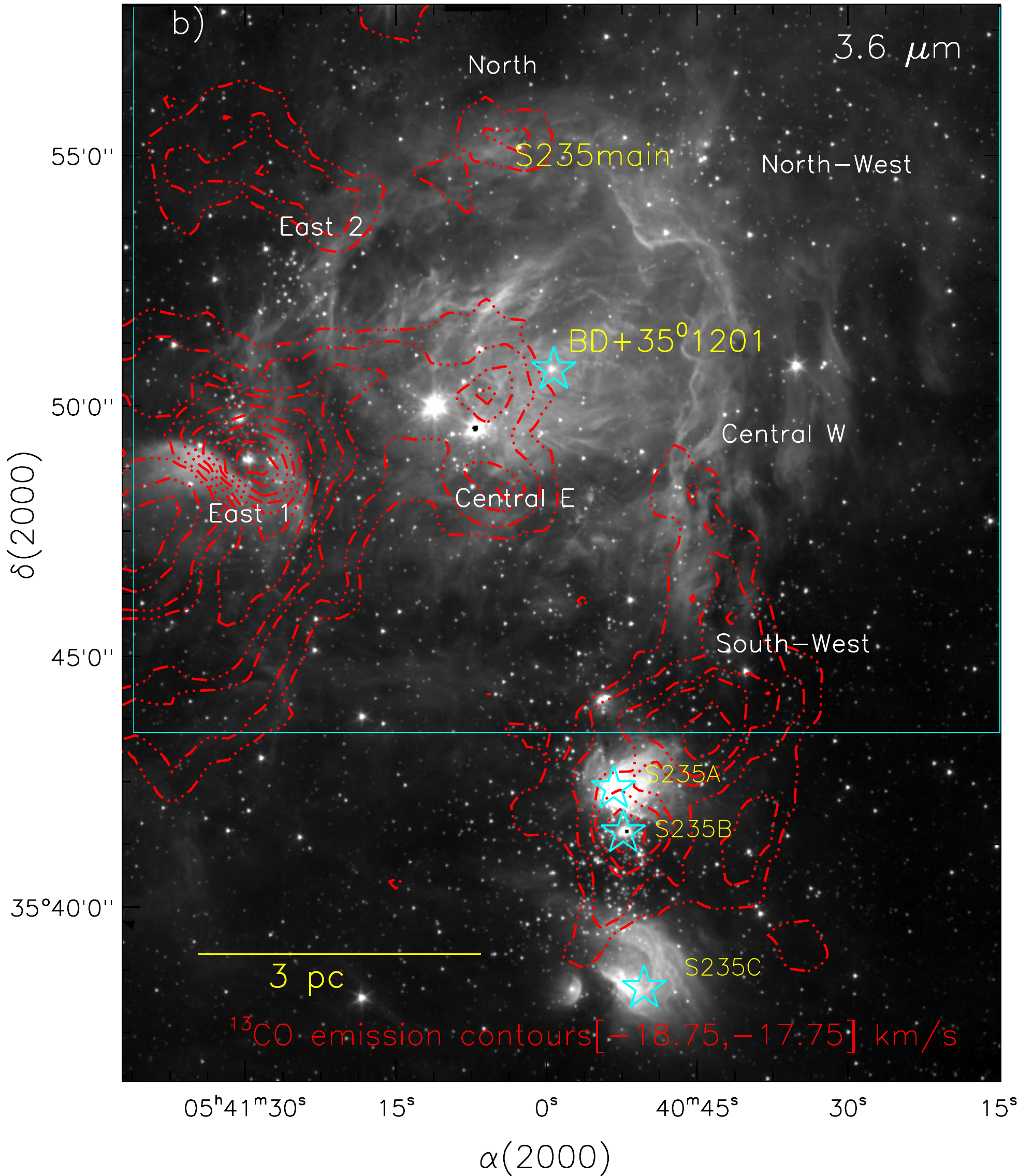}
\caption{\scriptsize a) {\it Spitzer} 3.6 $\mu$m image is overlaid with the $^{12}$CO emission integrated over $-$18.75 to $-$17.75 km s$^{-1}$. 
The $^{12}$CO contours are shown with levels of 
35.933 K km s$^{-1}$ $\times$ (0.2, 0.3, 0.45, 0.5, 0.6, 0.7, 0.8, 0.9, and 0.98). 
b) {\it Spitzer} 3.6 $\mu$m image is overlaid with the $^{13}$CO emission integrated over $-$18.75 to $-$17.75 km s$^{-1}$. 
The $^{13}$CO contours are shown with levels of 
14.402 K km s$^{-1}$ $\times$ (0.09, 0.15, 0.25, 0.3, 0.4, 0.5, 0.6, 0.7, 0.8, 0.9, and 0.98). 
In both the panels, different subregions are labeled \citep[also see Figure~1 in][]{dewangan16}. 
The scale bar on the bottom left shows a size of 3 pc at a distance of 1.8 kpc in each panel. 
In each panel, other marked symbols and labels are similar to those shown in Figure~\ref{fig6}a.}
\label{fg7x} 
\end{figure*}

In Figures~\ref{fig6} and~\ref{fig7}, we present the $^{12}$CO and $^{13}$CO emission contours overlaid on the 
{\it Spitzer} 3.6 $\mu$m image, respectively. 
Figures~\ref{fig6}a and~\ref{fig7}a show the {\it Spitzer} 3.6 $\mu$m image overlaid with the $^{12}$CO and 
$^{13}$CO emission integrated over $-$24 to $-$18 km s$^{-1}$, respectively. 
Figures~\ref{fig6}b and~\ref{fig7}b present the {\it Spitzer} 3.6 $\mu$m image superimposed with 
the $^{12}$CO and $^{13}$CO emission integrated over $-$18 to $-$13.5 km s$^{-1}$, respectively. 
In Figures~\ref{fg7x}a and~\ref{fg7x}b, we show the $^{12}$CO and $^{13}$CO emission contours overlaid on the {\it Spitzer} 3.6 $\mu$m image, respectively. These emissions are integrated over $-$18.75 to $-$17.75 km s$^{-1}$, which is an intermediate velocity range between two velocity 
peaks (see Figure~\ref{fig4}b). 
In Figures~\ref{fig6},~\ref{fig7}, and~\ref{fg7x}, we have highlighted previously known different subregions 
(i.e. East~1, East~2, North, North-West, Central W, Central E, South-West, S235A, S235B, and S235C) 
and the positions of the ionizing sources within E-S235. 

Based on Figures~\ref{fig4}c,~\ref{fig5}c,~\ref{fig6},~\ref{fig7}, and~\ref{fg7x}, the sites, ``S235ABC", East~1, and South-West subregions are 
spatially found at the junction of the molecular clouds within E-S235 (also see Section~\ref{sec:disc} for more details). 
\subsection{Dust continuum clumps and ionized emission in E-S235}
\label{subsec:dstmas}
\begin{figure*}
\epsscale{0.52}
\plotone{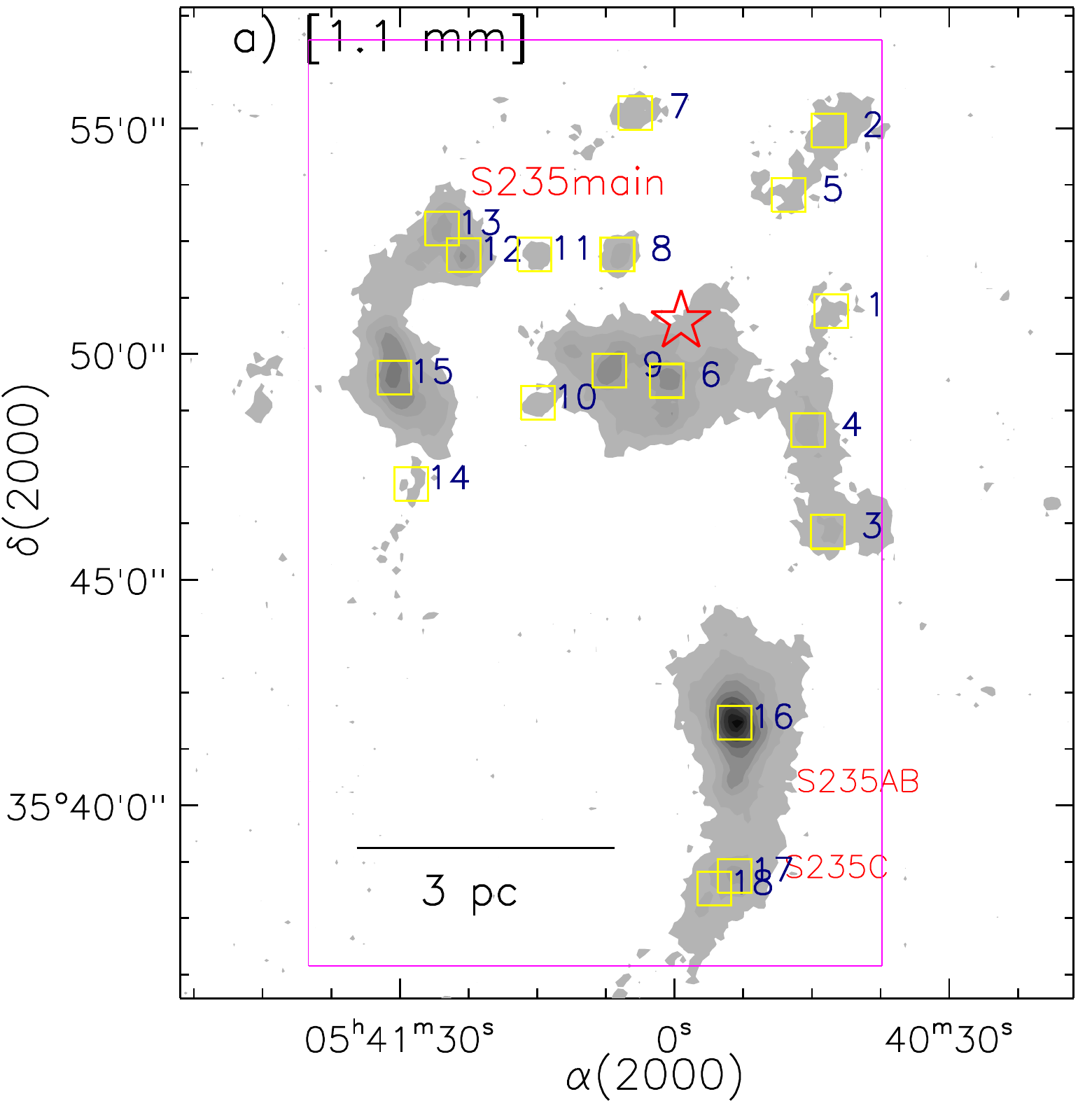}
\epsscale{0.45}
\plotone{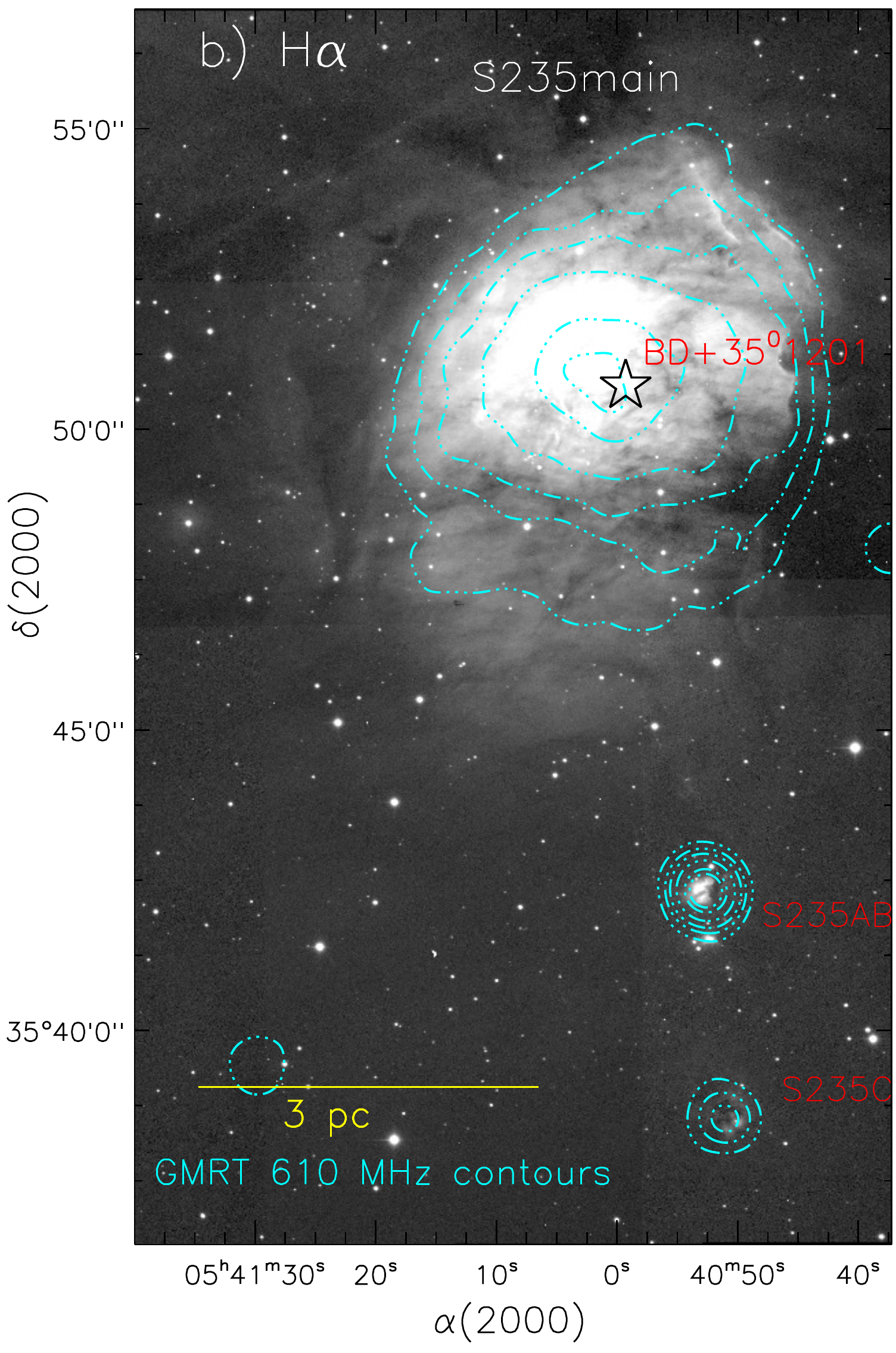}
\caption{\scriptsize a) Bolocam 1.1 mm dust continuum contour map (beam size $\sim$33$\arcsec$) of E-S235.
The contour levels are 2.5, 8, 15, 20, 30, 40, 55, 70, 85, and 95\% of the peak value i.e. 3.39 Jy/beam.
The positions of dust clumps at 1.1 mm are highlighted by yellow squares and are also labeled.
The clump nos 1--15 are distributed toward the ``S235main" (also see Paper~II), while the clump nos 16--18 are found toward the ``S235ABC".
The solid magenta box encompasses the area shown in Figure~\ref{fig8}b. 
b) IPHAS H$\alpha$ gray-scale image is overlaid with the GMRT 610 MHz emission contours. 
The GMRT 610 MHz dotted-dashed contours (in cyan) are shown with levels of 0.01, 0.024, 0.04, 0.07, 0.12, and 0.15 Jy/beam. 
In both the panels, other marked symbols and labels are similar to those shown in Figure~\ref{fig6}a.}
\label{fig8}
\end{figure*}

In this section, we study the dust continuum clumps and the distribution of ionized emission in E-S235. 
The dust continuum map at 1.1 mm allows us to depict the dense and cold regions. 
Figure~\ref{fig8}a shows the gray-scale Bolocam dust continuum map at 1.1 mm overlaid with the dust continuum clumps at 1.1 mm. 
These clumps were obtained from the Bolocam source catalog v2.1 \citep{ginsburg13}.
For our selected target region, we find eighteen clumps and fifteen out of them (nos 1-15; see Figure~\ref{fig8}a) have been reported in Paper~II. 
These fifteen clumps (i.e. nos 1--15) are distributed toward the ``S235main", while the remaining three clumps (nos 16, 17, and 18) are seen toward the ``S235ABC".
In Paper~II, the clump masses of first fifteen clumps have been computed which vary between 7 M$_{\odot}$ and 285 M$_{\odot}$ (see Table~1 in Paper~II).
The most massive dust clump (i.e. no. 15; M$_{clump}$ $\sim$285 M$_{\odot}$) is associated with the East~1 subregion.
The mass of a clump seen in the site South-West is reported to be about 60 M$_{\odot}$ (see ID \#3 in Table~1 in Paper~II).
Following Paper~II, we have also computed the clump masses of three clumps nos 16, 17, and 18 
(see Equation~1 in Paper~II for M$_{clump}$), 
which are $\sim$635, 70, and 40 M$_{\odot}$, respectively. 
In the calculation, we have used a dust absorption coefficient ($\kappa_\nu$ =) 1.14\,cm$^2$\,g$^{-1}$ \citep[e.g.][]{enoch08,bally10},  
a distance ($D$ =) 1.8 kpc, and a dust temperature ($T_d$ =) 20 K \citep[e.g.][]{kirsanova14}. 
A dust clump (i.e. no. 16; M$_{clump}$ $\sim$635 M$_{\odot}$) is seen toward the sites S235A and S235B (hereafter ``S235AB"). 
Two clumps (nos 17 and 18; M$_{clump}$ $\sim$40-70 M$_{\odot}$) are found toward the S235C region.

Figure~\ref{fig8}b shows an H$\alpha$ image overlaid with the GMRT 610 MHz emission. 
Both the images trace the distribution of ionized emission in E-S235. 
Spitzer-IRAC images also show extended bright features in E-S235 and each feature contains the radio continuum emission at its interior (see Figures~\ref{fig6}a and~\ref{fig8}b).
In Figure~\ref{fig8}b, we find the spatial match between the H$\alpha$ emission and the radio continuum emission. 
The ionized emission is found toward the sites ``S235main", S235A, S235B, and S235C.
The ``S235main" is powered by a single massive star BD+35$^{o}$1201 of O9.5V type, while the sites, S235A, S235B, and S235C are 
linked with the compact H\,{\sc ii} regions, which are mainly excited by B1V--B0.5V stars \citep[see Table~1 in][]{bieging16}. 
The positions of the ionizing sources are marked in Figure~\ref{fig6}a.
These results are also in agreement with estimations of Lyman continuum photons using the GMRT 610 MHz data.
We have also used the GMRT 610 MHz data to compute the dynamical ages (t$_{dyn}$) of the H\,{\sc ii} regions associated with ``S235AB" and S235C sites. 
Adopting a typical value of n$_{0}$ (as 10$^{3}$(10$^{4}$) cm$^{-3}$), we estimated the dynamical ages of these H\,{\sc ii} regions to be $\sim$0.06(0.22) Myr. 
In Paper~II, the dynamical age (t$_{dyn}$) of the H\,{\sc ii} region linked with ``S235main" was reported to be $\sim$1 Myr.
Based on these results, it appears that the H\,{\sc ii} regions associated with the ``S235AB" and S235C sites are relatively younger than 
the H\,{\sc ii} region linked with ``S235main". 
\subsection{Young stellar populations and their clustering}
\label{subsec:phot1}
\begin{figure*}
\epsscale{0.55}
\plotone{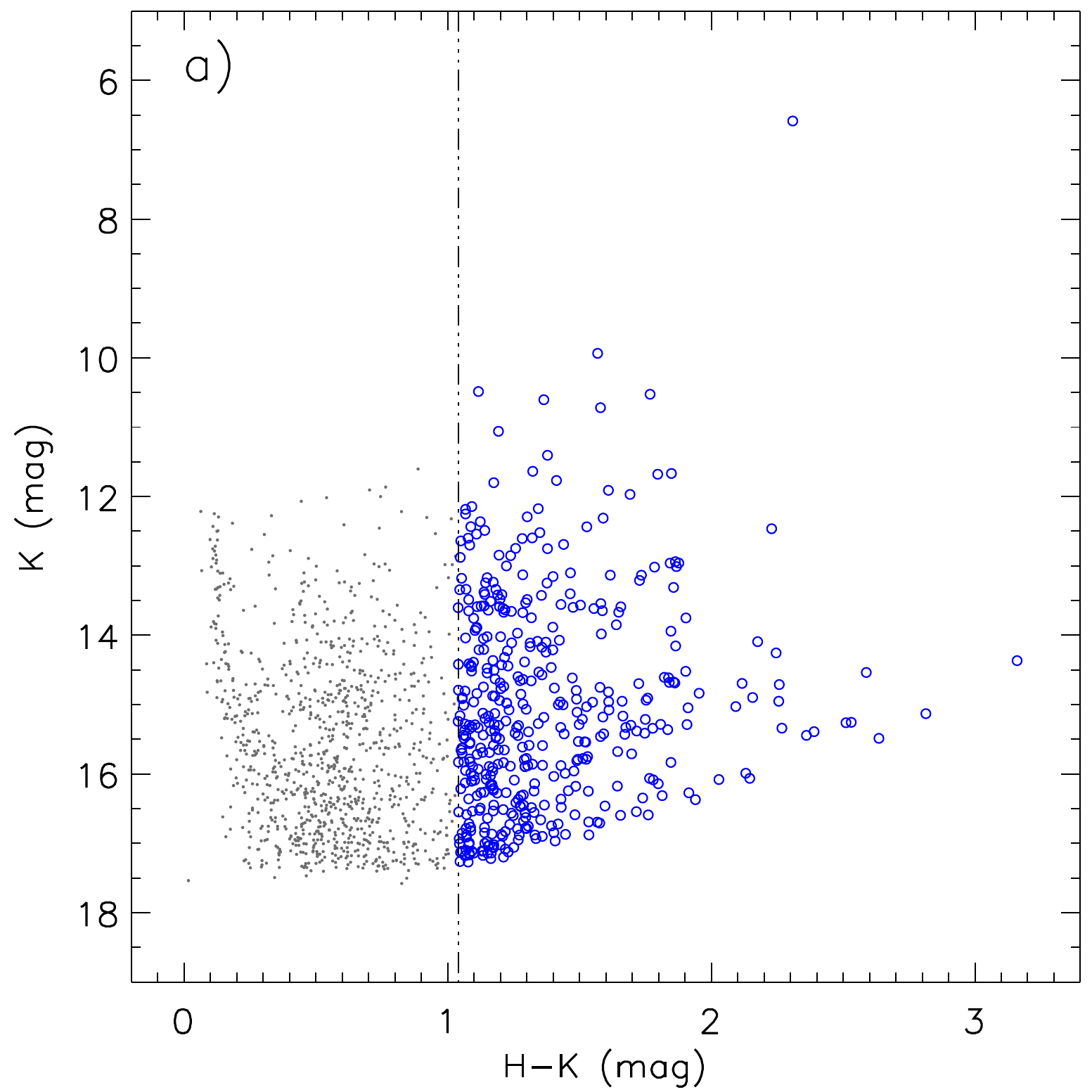}
\epsscale{0.59}
\plotone{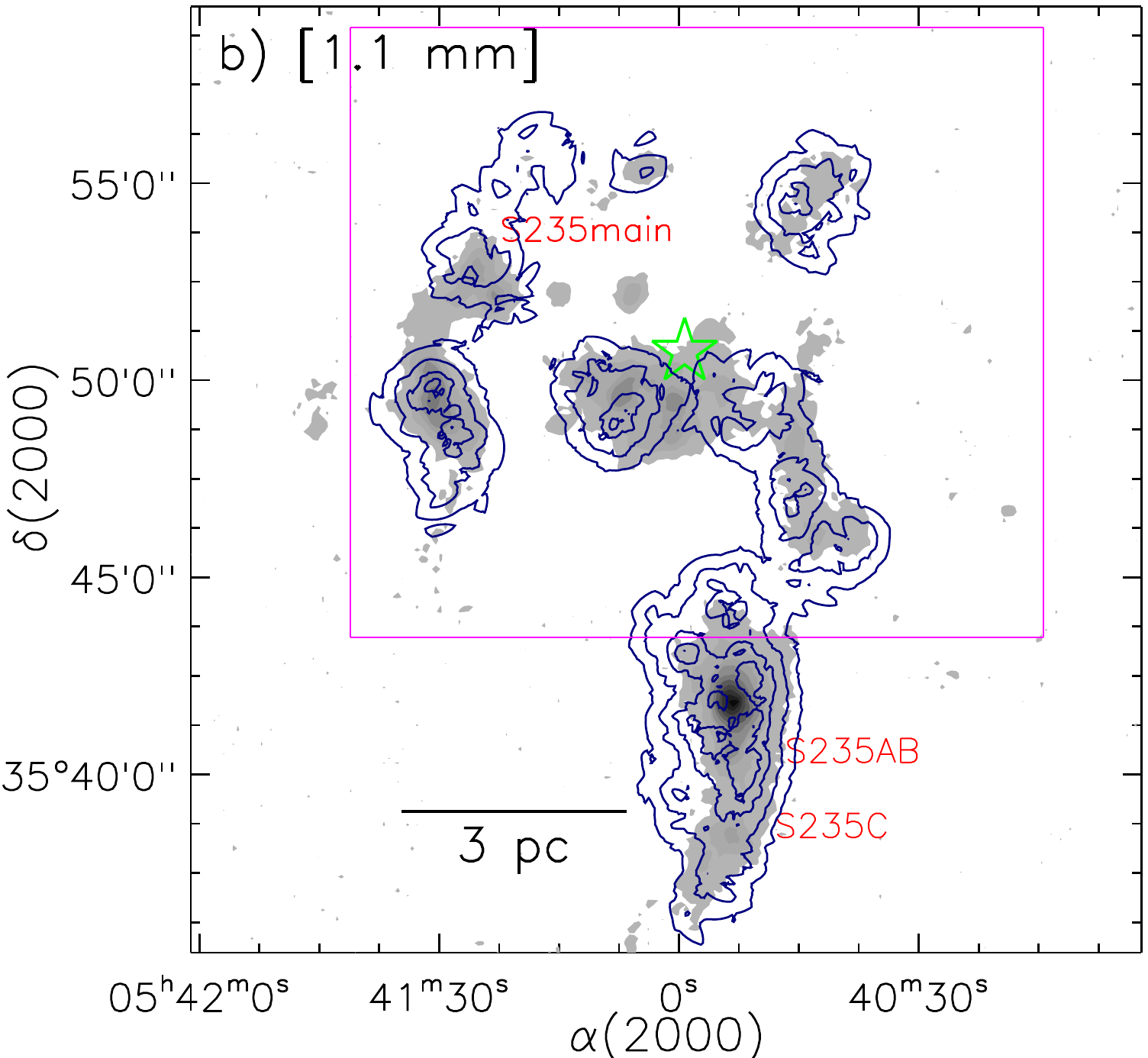}
\caption{\scriptsize 
a) Selection of embedded young stellar populations within E-S235. 
The GPS-2MASS color-magnitude diagram (H$-$K/K) of the sources detected in H and K bands.
The identified YSOs are marked by open blue circles, while the dots in gray color refer the stars with only photospheric emissions. 
In the NIR H$-$K/K plot, we have plotted only 1001 out of 5401 stars with photospheric emissions. 
Due to large numbers of stars with photospheric emissions, only some of these stars are randomly shown in the NIR H$-$K/K plot. 
A dotted-dashed line separates the identified YSOs against the stars with photospheric emissions. 
b) The surface density contours (in navy) of all the identified YSOs are overlaid on the 1.1 mm dust continuum map.
The background map is similar to the one shown in Figure~\ref{fig8}a. 
The contours are shown at 5, 10, 25, and 60 YSOs/pc$^{2}$, from the outer to the inner side. 
A solid box shows the area investigated in Paper~II. Other marked symbol and labels are similar to those shown in Figure~\ref{fig6}a.}
\label{fig9}
\end{figure*}

In this section, to probe the star formation activity, we have carried out an investigation of young stellar objects (YSOs) in our selected region.
The YSOs are identified using their infrared excess emission which are produced by the envelope and/or the disk of dust around them.
Previously, many authors identified the young stellar populations in our selected target \citep[e.g.,][]{allen05, kirsanova08,dewangan11,camargo11,chavarria14,kirsanova14,bieging16,dewangan16}. 
In the present work, we have employed the UKIDSS-GPS NIR data for depicting more deeply embedded and faint young stellar populations.
It is possible because the UKIDSS-GPS NIR data are three magnitudes deeper than 2MASS. 
In Paper~II, the GPS NIR data were mainly employed for the ``S235main" site.
Following the analysis presented in Paper~II, we have selected sources that have detections only in the  H and K bands. 
To identify infrared excess sources, we have utilized a color-magnitude (H$-$K/K) diagram (see Figure~\ref{fig9}a). 
The diagram allows to select embedded sources having H$-$K $>$ 1.04. 
This color criterion is chosen based on the color-magnitude analysis of a nearby control field (size $\sim$17$\farcm$6  $\times$ 15$\farcm$5; central coordinates: $\alpha_{J2000}$ = 05$^{h}$43$^{m}$35$^{s}$.8, $\delta_{J2000}$ = +35$\degr$20$\arcmin$20$\arcsec$.4). 
This condition yields 440 deeply embedded infrared excess sources. 

\begin{figure*}
\epsscale{1}
\plotone{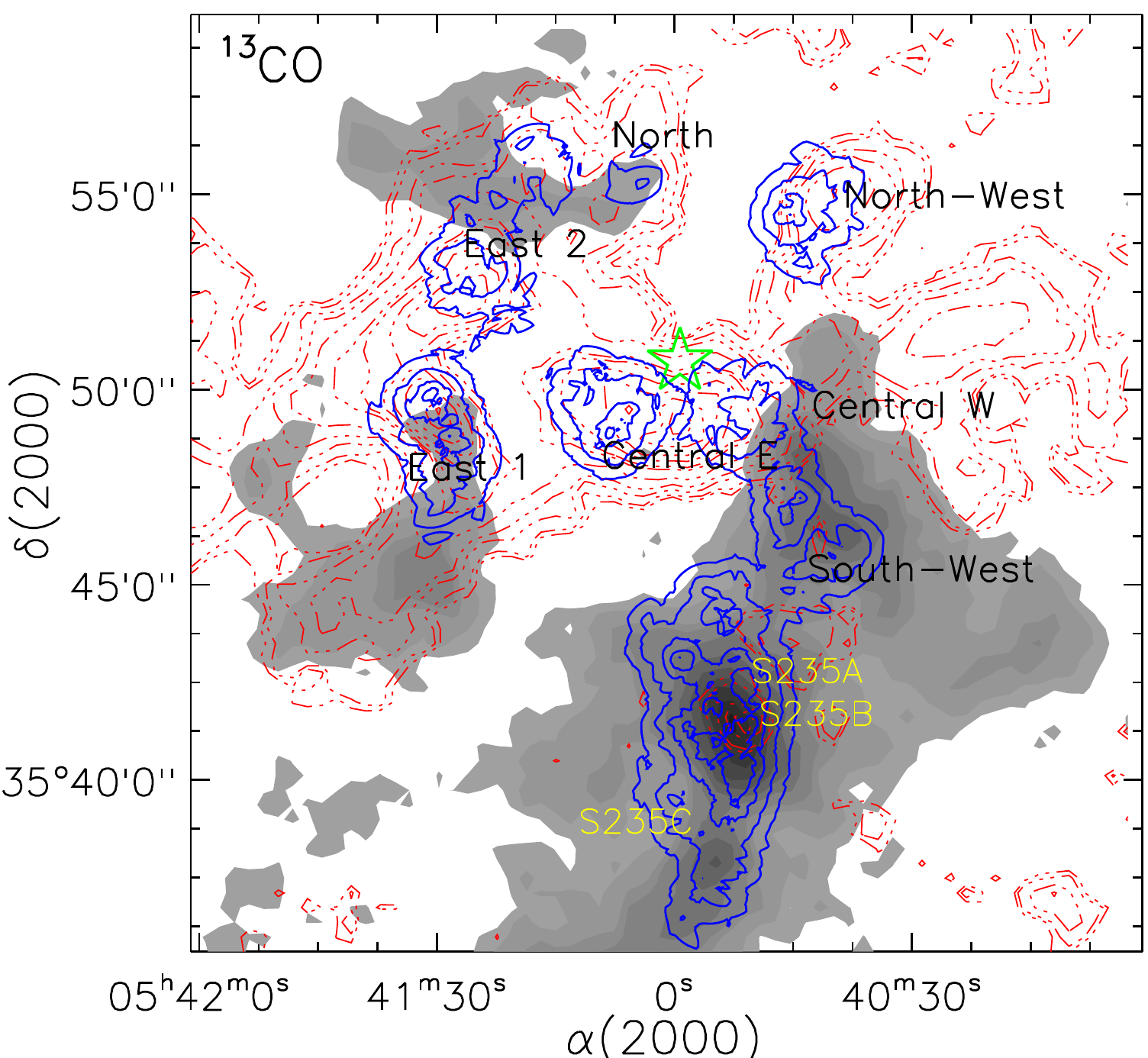}
\caption{\scriptsize The surface density contours (in blue) of all the identified YSOs are overlaid on the molecular $^{13}$CO maps.
The background maps are similar to the one shown in Figure~\ref{fig5}c. 
The surface density contours are shown at 5, 10, 25, and 60 YSOs/pc$^{2}$, from the outer to the inner side. 
Other marked symbol and labels are similar to those shown in Figure~\ref{fig6}a.}
\label{fig10}
\end{figure*}

To examine the individual groups or clusters of YSOs in E-S235, the surface density map of YSOs is produced using the 
nearest-neighbour (NN) technique \citep[also see][for more details]{gutermuth09,bressert10}. 
We have generated the surface density map of all the selected 440 YSOs, in a manner similar to that utilized in Paper~I. 
The map was obtained using a 5$\arcsec$ grid and 6 NN at a distance of 1.8 kpc. 
Figure~\ref{fig9}b shows the resultant surface density contours of YSOs overlaid on the 
dust continuum map at 1.1 mm. The contour levels are shown at 5, 10, 25, and 60 YSOs/pc$^{2}$, 
increasing from the outer to the inner regions. 
The figure clearly reveals the spatial correlation between YSOs surface density and dense clumps. 
The clusters of YSOs are spatially found toward all the subregions/sites, 
East~1, East~2, North, North-West, Central W, Central E, South-West, S235A, S235B, and S235C (see Figure~\ref{fig9}b). 
In E-S235, this outcome appears in agreement with the work of \citet{chavarria14} \citep[also see Figure~21 in][]{bieging16}. 
In Figure~\ref{fig10}, we have also overlaid the surface density contours on the molecular $^{13}$CO emission maps of the redshifted 
and the blueshifted molecular components. This figure helps us to trace the clustering of YSOs at the junction of two molecular clouds, indicating the intense star formation activities.
\section{Discussion}
\label{sec:disc}
Our present work expands upon the outcomes of Paper~II with a more detailed multi-wavelength analysis 
of E-S235 containing the ``S235main" and ``S235ABC" sites. In Paper~II, the analysis was restricted mainly 
to the ``S235main" site and it was suggested that the star formation activities toward the five subregions 
(i.e. East~2, North, North-West, Central~W, and Central~E) could be explained by the expanding H\,{\sc ii} 
region via ``collect and collapse". It was also reported that the clusters of YSOs in East~1 subregion might be 
originated by compression of the pre-existing dense material by the expanding H\,{\sc ii} region.
Furthermore, it was argued that the CCC process might be acting in E-S235.
Using the hydrogen radio recombination line observations, \citet{balser11} and \citet{anderson15} reported the velocity of the ionized gas to be 
about $-$25.6 and $-$15.3 km s$^{-1}$ in the H\,{\sc ii} regions linked 
with the ``S235main" and ``S235ABC", respectively. 
Considering the velocities of the molecular and ionized gases in E-S235, 
it is obvious that the molecular cloud linked with the sites ``S235ABC" is redshifted relative to 
the cloud associated with the site ``S235main" (also see Figures~\ref{fig6} and~\ref{fig7}). 
Hence, it appears that the molecular cloud containing ``S235ABC" is located at the rear side with respect to 
the ``S235main" site.
It is in agreement with the CCC scenario. 
However, a detailed study of the CCC process in E-S235 is still lacking. 
Hence, considering this interesting possibility, our present analysis has been focused to get more insights of the CCC in E-S235.
The CCC process has also been demonstrated in the numerical simulations 
\citep[see][and references therein]{habe92,inoue13,takahira14,haworth15a,haworth15b}. 
As mentioned before, the signposts of the CCC are the bridging feature connecting the two clouds in velocity, the complementary distribution of the two 
colliding clouds, and broad CO line wing in the intersection of the two clouds. 
The presence of a broad bridge feature in the velocity space can be interpreted as an evidence of a compressed layer of gas due to the collision 
between the clouds seen along the line of sight \citep[e.g.,][]{haworth15a,haworth15b,torii17,torii17b,fukui17}. 
One can also note that the synthetic observations presented 
in \citet{haworth15a,haworth15b} were made with an observer viewing angle parallel to the collisional axis so that the two colliding clouds are spatially coincident along the line of sight. 
In such a case, two velocity peaks separated by intermediate intensity emissions are found in the velocity space. The same is the case in M20 studied by \citet{torii17}. 
If one observes a CCC with a viewing angle inclined relative to the collisional axis, or if a CCC is an offset collision, the observed position-velocity diagrams and line profiles may be different from 
those presented in \citet{haworth15a,haworth15b} and \citet{torii17} \citep[also see][]{fukui17}. 
Using the $^{12}$CO (J=1--0) and $^{13}$CO (J=1--0) line data, two molecular clouds are observed in the direction of E-S235 
and are interconnected in space as well as in velocity (see Section~\ref{sec:coem}). 
The molecular second momentum map has revealed the colliding interface or the boundaries of the molecular clouds (see Figure~\ref{fg5x}c). 
A possible complementary molecular pair at [$-$21, $-$20] km s$^{-1}$ and [$-$16, $-$15] km s$^{-1}$ is also found in E-S235. 
Furthermore, the existence of the broad bridge feature (see Figures~\ref{fig4}c and~\ref{fig5}c)  and its flattened molecular profiles (see Figure~\ref{fg6x}) indicate 
the onset of the CCC process and the presence of the turbulent gas excited by this mechanism. 
Taken together, all the observed signatures are in agreement with the outcomes of models of the CCC \citep[e.g.,][]{inoue13,takahira14,haworth15a,haworth15b}.

In E-S235, we find sites, East~1, ``S235ABC", and South-West seen in the spatially overlapped zones of two molecular clouds (see Figures~\ref{fg5x}--\ref{fg6x} and also Section~\ref{sec:coem}). 
In the sites of the CCC, one may expect the higher values of ratio of CO 3-2/1-0 toward the colliding interface of the colliding clouds \citep{torii15}.
\citet{bieging16} studied E-S235 in CO (2--1), $^{13}$CO (2--1), and CO (3--2) and presented 
velocity channel maps of CO (J = 3--2)/(J = 2 -1) intensity ratios, 
map of velocity dispersion for $^{13}$CO (2-1), and distribution of CO and $^{13}$CO (2-1) 
excitation temperature derived by an LTE model. Using these published results toward 
the colliding interfaces (i.e. East~1, ``S235ABC", and South-West) in E-S235, 
the values of CO (J = 3--2)/(J = 2 -1) intensity ratios, velocity dispersion for $^{13}$CO (2-1), 
and excitation temperature are found to be about 0.6--0.9, 2--4 km s$^{-1}$, and 25--40 K, respectively.
These values are noticeably high toward the sites, East~1, ``S235ABC", and South-West, and 
may be explained by the CCC \citep[e.g.][]{torii15,torii17,torii17b} (also see Figure~\ref{fg5x} 
for the molecular first and second momentum maps in this paper). 
All these sites are associated with intense star formation activities. 
The sites, S235A and S235B contain compact H\,{\sc ii} regions excited by B-type stars.
The site East~1 is not associated with any ionized emission and contains many embedded protostars 
that show outflow activities (see Papers I and II). A massive clump (i.e. M$_{clump}$ $\sim$285 M$_{\odot}$) 
is found toward the East~1. Additionally, this subregion is considered as the youngest star-forming 
site in the S235main \citep[see][]{kirsanova14}. 
In E-S235, the most massive clump (i.e. M$_{clump}$ $\sim$635 M$_{\odot}$) is observed 
toward the sites ``S235AB" that have significant gas mass within $\sim$2 pc of the clusters.
Considering the velocity separation (i.e. $\sim$4 km s$^{-1}$) of the two colliding clouds in the E-S235, 
a typical collision timescale is computed to be $\sim$0.5 Myr.
The dynamical ages of the H\,{\sc ii} regions associated with the sites ``S235AB" are found to be 0.06--0.22 Myr. 
An average age of the Class~I and Class~II YSOs is estimated to be $\sim$0.44 Myr and $\sim$1--3 Myr, respectively \citep{evans09}. 
Taking into account these timescales, 
it seems that the formation of youngest populations and massive stars in the interface of two clouds 
(i.e. East~1, ``S235ABC", and South-West) might be influenced by the CCC about 0.5 Myr ago. 
However, the stellar populations older than the collision timescale might have formed prior to the collisions of clouds. 
Hence, we cannot rule out the onset of star formation activities before the collision in E-S235.

Considering the results presented in Paper~II and the outcomes of this paper, it reinforces the idea that 
the E-S235 is a very promising triggered star formation site where 
the ``collect and collapse" \citep[e.g.][]{kang12,kirsanova14,chavarria14,bieging16} and 
the CCC processes seem to have influenced the star formation activities. 
\section{Summary and Conclusions}
\label{sec:conc}
In this paper, we have carried out an extensive study of E-S235 (harboring the ``S235main" and ``S235ABC" sites) 
using the multi-wavelength data.
The major results of our multi-wavelength analysis are the following:\\
$\bullet$ Using the $^{12}$CO (J=1--0) and $^{13}$CO (J=1--0) line data, the molecular gas emission in the direction of E-S235 
reveals two noticeable molecular clouds (having velocity 
peaks at $\sim$$-$20.5 and $\sim$$-$16.5 km s$^{-1}$), which are separated by $\sim$4 km s$^{-1}$ in the velocity space. The boundaries of the two clouds (or the colliding interface) are traced using the molecular momentum maps. 
The redshifted molecular component ($-$18 to $-$13 km~s$^{-1}$) is linked with 
the sites S235A, S235B, and S235C, while the blueshifted one ($-$24 to $-$18 km~s$^{-1}$) 
contains the site ``S235main".\\
$\bullet$ The position-velocity space of $^{12}$CO and $^{13}$CO depicts a broad bridge feature. 
In this configuration, the redshifted and the blueshifted molecular components are interconnected by a lower 
intensity intermediate velocity emission.\\ 
$\bullet$ In the velocity channel maps of $^{12}$CO and $^{13}$CO, a possible complementary 
molecular pair at [$-$21, $-$20] km s$^{-1}$ and [$-$16, $-$15] km s$^{-1}$ is also traced.\\
$\bullet$ The sites, East~1, ``S235ABC", and South-West are found in the spatially overlapped zones of two molecular clouds.\\
$\bullet$ Deep UKIDSS NIR photometric analysis of point-like sources gives a total of 440 YSOs in E-S235 and depicts significant clustering of 
young stellar populations toward the sites located in the interface, and the ``S235main".\\ 
$\bullet$ The sites, ``S235ABC" harbor young compact H\,{\sc ii} regions having dynamical ages of $\sim$0.06--0.22 Myr.\\ 
$\bullet$ The sites, ``S235ABC", South-West, and East~1 contain dust clumps (having M$_{clump}$ $\sim$40 to 635 M$_{\odot}$). \\
$\bullet$ A typical collision timescale in E-S235 is estimated to be $\sim$0.5 Myr.

With the observational outcomes presented in this paper, we conclude that the CCC process might have influenced the star formation activities (including massive stars) at the junction about 0.5 Myr ago. 
\acknowledgments
We thank the anonymous reviewers for constructive comments and suggestions which improved the scientific content of this paper. 
The research work at Physical Research Laboratory is funded by the Department of Space, Government of India. 
This work is based on data obtained as part of the UKIRT Infrared Deep Sky Survey. This publication 
made use of data products from the Two Micron All Sky Survey (a joint project of 
the University of Massachusetts and the Infrared Processing and Analysis Center / California Institute of 
Technology, funded by NASA and NSF), archival data obtained with the {\it Spitzer} 
Space Telescope (operated by the Jet Propulsion Laboratory, California Institute 
of Technology under a contract with NASA). 
The Canadian Galactic Plane Survey (CGPS) is a Canadian project with international partners. 
The Dominion Radio Astrophysical Observatory is operated as a national facility by the 
National Research Council of Canada. The Five College Radio Astronomy Observatory 
CO Survey of the Outer Galaxy was supported by NSF grant AST 94-20159. The CGPS is 
supported by a grant from the Natural Sciences and Engineering Research Council of Canada. 
The GMRT is run by the National Centre for  Radio Astrophysics of the Tata Institute of
Fundamental Research. 
This paper makes use of data obtained as part of the INT Photometric H$\alpha$ 
Survey of the Northern Galactic Plane (IPHAS, www.iphas.org) carried out 
at the Isaac Newton Telescope (INT). The INT is operated on the 
island of La Palma by the Isaac Newton Group in the Spanish Observatorio del Roque de los Muchachos of 
the Instituto de Astrofisica de Canarias. The IPHAS data are processed by the Cambridge Astronomical Survey 
Unit, at the Institute of Astronomy in Cambridge. 

\end{document}